\documentclass[lineno]{aastex631}
\usepackage{amsmath,amssymb,CJK,textcomp}

\newcommand{\R}{\textcolor{red}}
\newcommand{\B}{\textcolor{blue}}

\journalinfo{The Astrophysical Journal (preprint)} 
\shortauthors{Battams et al.}

\accepted{2025 March 30}

\shorttitle{Geminid WISPR models}
\shortauthors{Battams et al.}

\graphicspath{{./}{figures/}}

\begin{document}
\begin{CJK*}{UTF8}{gbsn}
\title{A Comparison of Geminid Models with the PSP/WISPR-observed Phaethon Dust Trail}

\correspondingauthor{Karl Battams}
\email{karl.battams.civ@us.navy.mil}

\author[0000-0002-8692-6925]{Karl Battams}
\affil{US Naval Research Laboratory, 4555 Overlook Avenue SW, Washington, DC 20375, USA}

\author[0000-0002-1242-1457]{Galina O. Ryabova}
\affil{Tomsk State University, Tomsk, Russian Federation}

\author[0000-0002-6420-8404]{Angel J. Gutarra-Leon}
\affiliation{George Mason University, Fairfax, VA, USA}

\author[0000-0003-2685-9801]{Jamey R. Szalay}
\affil{Princeton University, Princeton, NJ 08544, USA}

\author[0000-0002-8353-5865]{Brendan M. Gallagher}
\affil{US Naval Research Laboratory, 4555 Overlook Avenue SW, Washington, DC 20375, USA}

\author[0000-0002-8658-3811]{Wolf Cukier}
\affil{University of Chicago, Chicago, IL, USA}

\begin{abstract}
White-light observations from the WISPR instrument on NASA's Parker Solar Probe recently revealed the presence of a narrow, dense dust trail close to the orbit of asteroid 3200 Phaethon. Although Geminid-related, it aligns imperfectly with Phaethon's orbit and known Geminid meteoroid orbits. To address the nature of this dust trail, we performed a detailed comparison between the WISPR trail observations and several well-developed Geminid models. Simulating these models in the WISPR field of view visually demonstrates that the WISPR trail almost certainly represents the true ``density core'' of the Geminid stream. Trends in model trail width and offset from Phaethon's orbit, both as functions of true anomaly, agree with observations to varying extents. All the models, however, place their apparent core interior to the parent orbit due to Poynting-Robertson forces, contradictory to the WISPR trail which is exterior to Phaethon's orbit. Therefore, Phaethon's current orbit likely does not represent the orbit of the system parent, which most probably had a larger semi-major axis. These findings provide new initial conditions for future Geminid models, with WISPR identifying the Geminid core's position.

\end{abstract}


\section{Introduction} \label{sec:intro}

As one of Earth's most prolific meteor showers, the Geminids are a continual source of intrigue for astronomers and the general public, reaching their peak in mid-December each year and producing up to 120 meteors per hour \citep{2019pimo.conf..114R}. Studies of this meteoroid stream were energized by the discovery of the apparently asteroidal near-Sun object 3200 Phaethon as 1983TB \citep{IAUC3878}, which was identified as the assumed Geminid parent \citep{Whipple1983} soon after discovery. With such an abundant meteoroid stream associated with an asteroid, versus the cometary origin of most other streams (see \citealt{2022arXiv220301397J, 2022arXiv220910654Y} and references therein), the Phaethon--Geminid system has been the target of ongoing scrutiny as the community seeks to understand the processes that led to its formation and evolution. 

The first reliable observations of the Geminids date back to 1862 \citep{1926MNRAS..86..638K}, but detailed attempts at modeling the origins and evolution of the stream did not occur until the 1980s, with studies by \citet{Fox1982, Fox1983}, \citet{1985AVest..19..152L}, \citet{1985MNRAS.217..523J}, \citet{Jones1986} attempting early computational models of the system. (A brief historical review was provided by \citealt{2014me13.conf..205R}.) These efforts, aided by Phaethon's relatively recent 1983 discovery, determined the origin of the Geminids likely dates back approximately 2000 years \citep{1999SoSyR..33..224R}. This estimate has been broadly adopted by much of the community, although some recent studies propose that this timeline could be substantially longer \citep[e.g.][]{Jo2024}. A comprehensive review of observational and theoretical studies of the stream was performed by \cite{2015CoSka..45...60N}. Further extensive reviews can be found related to the stream \citep{2019msme.book..161V} and its parent body \citep{2019msme.book..187K}, describing the progress in twenty-first-century research.

More recent Geminid modeling work has examined the following issues:
\citet{2020P&SS..19004965H} analyzed influence of observational errors on the heliocentric orbital elements using a simple model;
\citet{2021Icar..35413949M} estimated the particles masses that are excluded from the stream entirely;
\citet{2021MNRAS.507.4481R} modeled radiants for meteors of various masses;
\citet{2023A&A...673A.161C} used numerical models of the stream to draw and study chaos maps;
\citet{Jo2024} modeled the Geminid formation under the assumption that the particle's ejection happened via asteroid rotation; 
\citet{2022P&SS..21005378R} explored mean motion resonances in the Geminid stream; and 
\cite{Cukier2023} (a primary focus of this study) explored and modeled different Geminid formation mechanisms.

However, despite the fact that the number of publications related to the stream has long exceeded a thousand, many of the details of the formation and evolution of the stream have remained elusive. This is partly due to relatively limited direct observation of the stream, other than that which can be extracted from studying the Earth-encountering component of the stream, i.e. individual meteors. Since Phaethon's discovery, many ground-based surveys have unsuccessfully attempted to observe activity of this asteroid
\citep{1984Icar...59..296C, Green1985,1996Icar..119..173C, 2005ApJ...624.1093H, 2008Icar..194..843W}. Even within the past decade, ground-based surveys have sought signatures of small fragments \citep{Jewitt2018,Ye2018} and large dust particles \citep{Jewitt2019} without positive detection, despite uniquely favorable observing conditions.
 
A weak and short-term ($\sim$two day) perihelion activity of Phaethon was discovered in 2009, 2012 and 2016 \citep{2009IAUC.9054....3B, Jewitt2010, 2013AJ....145..154L, 2017AJ....153...23H} in near-Sun heliophysics imaging obtained by the NASA Solar Terrestrial Relations Observatory (STEREO). The discovery of this recurrent perihelion activity inspired a scientific discussion which has generated dozens of articles analyzing possible mechanisms of this activity, mass-loss rates, replenishment of the Geminid meteoroid stream, and so on. An important impetus for the discussion was that this activity was initially attributed to release of micron-sized dust \citep{2013ApJ...771L..36J}. However, \cite{Zhang2023} exploited more than twenty years of coronagraph and heliospheric imager observations from both STEREO and the ESA/NASA Solar and Heliospheric Observatory (SOHO), to demonstrate quite comprehensively that Phaethon's perihelion brightening is better attributed primarily, if not entirely, to sodium emission rather than dust. This discovery was supported by \cite{Hui2023} who also found a lack of Phaethon dust signatures in coronagraph observations from STEREO between 2008 and 2022.

The first direct observation of any kind of Geminid/Phaethon trail was reported by \citet{2014AJ....148..135A}, where the results of reprocessed Cosmic Background Explorer (COBE) Diffuse Infrared Background Experiment (DIRBE) thermal infrared survey data from space, obtained in 1989--1990, revealed a number of dust trails following orbits of several comets and 3200~Phaethon. However, other space-based thermal infrared observations dedicated to 3200 Phaethon and the possibly Geminid-related asteroids (155140) 2005 UD, and (225416) 1999~YC, found no evidence of their activity \citep{2022AJ....164..193K}.

In 2018, however, direct visible-light observations of the Geminid stream distant from Earth were finally obtained, and from the surprising source of the recently launched NASA Parker Solar Probe \citep[PSP,][]{Fox15} spacecraft. As detailed in \cite{Battams2020}, white-light images from PSP's Wide-Field Imager for Parker Solar Probe \citep[WISPR;][]{Vourlidas15} instrument revealed the presence of a faint dust trail appearing to perfectly follow the orbit of 3200 Phaethon. This work determined the visual (V) magnitude of the stream to be 15.8~$\pm$~0.3 per pixel, with a surface brightness of 25.0 mag arcsec$^{-2}$ as seen by PSP. This surface brightness value is further reduced to an estimated 30.8 at 1 au, and 31.7 at 2.3~au (aphelion), placing it below the detection capabilities of current ground-based facilities. Between 2020 and 2022, PSP's evolving orbit saw the spacecraft increasingly approach the orbit of Phaethon. As reported in \cite{Battams2022}, by August 2021 the minimum distance between PSP and Phaethon's orbit had fallen from 0.0765~au (in 2018) to 0.0277~au. This increasing proximity, combined with small changes in PSP's observing location, led to the discovery that the trail was in fact positioned slightly anti-sunward of Phaethon's orbit, with a small apparent separation between them that increased as a function of true anomaly (TA) along Phaethon's orbit (see Figure 6 of \citealt{Battams2022}). Hereafter, we will use ``TA'' throughout this manuscript to refer specifically to the true anomaly of Phaethon's orbit.

\begin{figure}[ht!]
\centering
\includegraphics[width=0.5\textwidth]{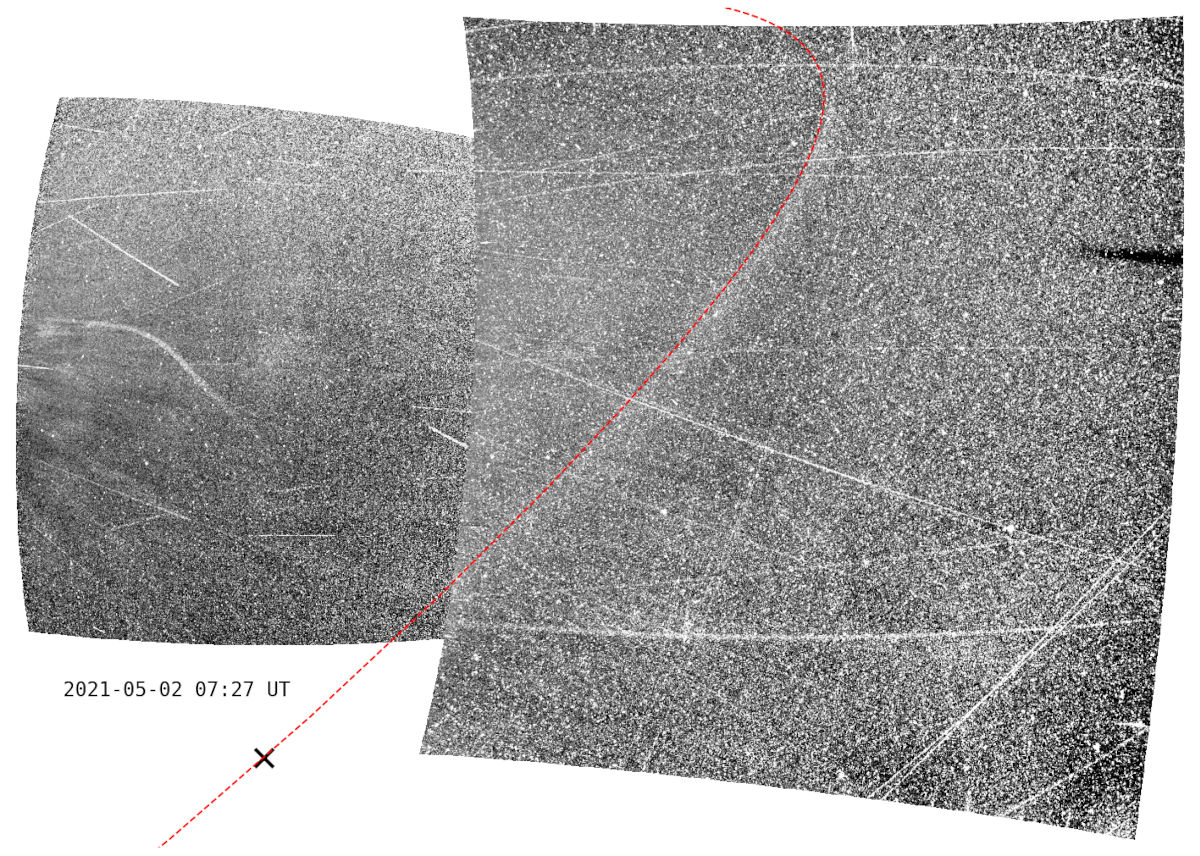}

\caption{WISPR-I and WISPR-O composite recorded on 2021 May 02 (Encounter 8). Phaethon's orbit is indicated by the dashed red line, with a black X indicating the closest point of Phaethon's orbit to \textit{PSP} at the time of observation, which happened to be outside of the field of view. This a replication of Figure 5 of \cite{Battams2022} }
\label{fig:phaethon_dust_image}
\end{figure}

Figure~\ref{fig:phaethon_dust_image} replicates Figure 5 of \cite{Battams2022}, presenting the clearest static image representation of the dust trail in the WISPR field of view (FOV). The trail manifests as a faint, diffuse structure immediately adjacent to the outside of Phaethon's orbit (red dashed line), not to be confused with the numerous individual, narrow dust streaks that cross the image. From some of PSP's viewing geometries, Phaethon's orbit overlaps more with the trail, but the trail is never seen interior to Phaethon's orbit, nor do we observe clear separation between the two. The photometry of this trail was shown to be entirely independent of both TA along the orbit, and of the position of Phaethon at the time. Thus, the dust observed by WISPR cannot have been recently ejected. These observations are important to keep in mind as we explore models in the later sections.

In this study we perform a direct comparison of the PSP/WISPR-observed dust trail with two different, fully-evolved models of the Geminids -- the Ryabova classical cometary model, and the Cukier-Szalay (hereafter, CS) model. The latter comprises three different formation scenarios, hereafter referred to individually as CS Basic, CS Cometary, and CS Violent, and collectively as the CS model. We explore these models both visually and in terms of apparent physical properties, focusing particularly on trends in observed versus modeled trail properties as a function of TA. The WISPR observations encompass more than 20 near-Sun encounters that represent six fundamentally different viewing geometries. For brevity, this study will focus on just three of these six geometries.

The purpose of this work is to first demonstrate that the trail observed by WISPR is indeed the Geminid stream, and that it exhibits the general properties we expect of such a stream. The secondary goal is to determine, discuss, and explain differences between models and observations to better inform any future modeling efforts. We will show that most of the models under consideration contain a centralized density that manifests as a narrow trail (core) in simulated WISPR observations, and mirror the properties observed in the actual WISPR observations. However, we will also show that models do not correctly predict the absolute placement of the Geminid core. The models place the bulk of the density inside of Phaethon's present orbit due to Poynting-Robertson forces. This differs fundamentally from WISPR observations, which show the trail marginally outside of Phaethon's orbit. Although this difference could result from some unidentified and complex physical process, we propose that the simpler and more intuitive explanation is that Phaethon does not follow the parent orbit, as first proposed by \cite{1985AVest..19..152L}).

It is non-trivial to compare models that assume different particle masses and densities, as these parameters directly affect other fundamental model properties. Thus, we emphasize here that the primary purpose of this work is not to evaluate the Ryabova and CS models with each other. While we do compare both to the WISPR observations, the Ryabova model is included primarily for the interpretation of Geminid shower features based on ground observations, whereas the CS model is more focused on comparison with WISPR observations. {\it A caveat}: While we opted to retain the respective author's naming of their models, we note that the fundamental assumptions of the Ryabova classical cometary model differ from those of the CS Cometary model. We have tried to avoid ambiguity in the resulting narrative, but the reader is advised to be aware of this unavoidable overlap in terminology. 

First we present a recap of the PSP mission, the WISPR instrument, and the observations. We then provide an overview of the models under consideration and the variable parameters of each. We also briefly describe ground-based observations from \cite{2003PAICz..91....1K} and \cite{ 2022A&A...667A.158B, 2022A&A...667A.157B}. We present both visual and numerical, trend-based comparisons between the WISPR observations and these models to assess the levels of agreement/disagreement between them, and evaluate the individual strengths of the model assumptions. We also provide a very brief update on the results presented in \cite{Battams2020} and \cite{Battams2022}.

\section{Data and Models} \label{sec:data}

\subsection{PSP/WISPR Overview}

PSP is a NASA heliophysics mission aimed broadly at exploring the extreme near-Sun environment. The mission seeks answers to several questions surrounding the transport of energy from the solar disk to the corona, exploring the dynamics and structure of magnetic fields in this region, and exploring the processes that lead to the transport and acceleration of energetic particles in this domain. PSP has already addressed several of these goals, and more, via a suite of instrumentation that includes both in-situ and remote sensing capabilities \citep{Raouafi2023}. 

PSP was launched in 2018 into an elliptical orbit around the Sun, and has utilized a series of Venus gravity-assist maneuvers to gradually decrease the spacecraft's perihelion distance from its first orbit value of 0.166 au (2018 November 5) to its current 0.046 au (December 2024). While the PSP instrumentation records observations throughout much of each orbit, the science observations are particularly focused on the short, $\sim$two-week ``Encounter'' periods that surround the spacecraft's perihelion passage. Throughout this manuscript, these Encounters are referred to by their Encounter number as, e.g., ``E6'' for Encounter six. 

As discussed in \cite{Battams2022}, these Encounters can be grouped according to their orbital similarity. That is, E1--E3, E4--E5, E6--E7, E8--E9, and E10--E16, E17--present, all followed near identical orbits with the aforementioned Venus fly-bys placing the spacecraft into each new orbital regime. Throughout all orbital regimes, the spacecraft reaches perihelion at approximately the same location in space relative to the Sun (i.e. the same longitude in a heliocentric coordinate system), but with differing perihelion distances. The pertinent consequence of this is that imaging instruments such as WISPR maintain approximately the same FOV (and same background star field) in every encounter, with only slight differences of effective FOV as the mission evolves. WISPR therefore also observes approximately the same portion of Phaethon's orbit during every encounter. Specifics of this can be found in \cite{Battams2020, Battams2022}. Table~\ref{tab:enc-overview} summarizes the orbit regimes, their date ranges, and the minimum distance of PSP to the Sun and Phaethon's orbit, during each regime to date.

\begin{table}[ht!]
\centering
\caption{Overview of the first twenty-one \textit{PSP}/WISPR solar encounter periods. The first columns lists the Encounter number range. The second column provides the start and end dates covered by those Encounters. The third column gives the perihelion distance of \textit{PSP} in astronomical units (au) during that orbit regime. The fourth column provides the minimum approach distance between \textit{PSP} and the orbit of 3200~Phaethon. \label{tab:enc-overview}}
\begin{tabular}{|c|c|c|c|}
\hline
\multicolumn{4}{|c|}{PSP/WISPR Encounters 1 -- 21 overview} \\
\hline
\textbf{ENC's} & \textbf{Date Range} & \textbf{\textit{PSP} Perihelion (au)} & \textbf{3200-PSP Min. Dist. (au)} \\
\hline
1 -- 3 & 2018 Oct 31 -- 2019 Sep 07 & 0.166 & 0.0765 \\
\hline
4 -- 5 & 2020 Jan 23 -- 2020 Jun 13 & 0.130 & 0.0563 \\
\hline
6 -- 7 & 2020 Sep 21 -- 2021 Jan 23 & 0.095 & 0.0387 \\
\hline
8 -- 9 & 2021 Apr 24 -- 2021 Aug 15 & 0.074 & 0.0277 \\
\hline
10 -- 16 & 2021 Nov 16 -- 2023 Jun 27 & 0.062 & 0.0136 \\
\hline
17 -- 21 & 2023 Sep 22 -- 2024 Oct 05 & 0.053 & 0.0025 \\

\hline
\end{tabular}
\end{table}

\subsubsection{WISPR Observations}\label{sec:wispr-obs}

WISPR comprises a pair of overlapping, white-light heliospheric imager-style cameras designed to observe solar outflows, such as the solar wind and coronal mass ejections, from very near the Sun and out into the inner heliosphere. It does this by placing the solar disk just outside of the FOV, and employing a series of light-rejecting baffles to screen out all direct sunlight, thus enabling the visual imaging of faint near-Sun structures. WISPR employs both an Inner (WISPR-I) and Outer (WISPR-O) camera, both of which use 1920$\times$2048 pixel Advanced Pixel Sensor (APS) detectors to obtain visible light images through 490--740 nm (WISPR-I) and 475--725 nm (WISPR-O) filters, at a plate scale of 1.2 arcmin/pixel, and with a combined effective FOV of 13.5$^{\circ}$ -- 53$^{\circ}$ (degrees elongation). During each Encounter, WISPR observations are recorded at varying data rates and exposure times, with the latter dictated by the region (brightness) of the corona visible in the FOV. This study focuses solely on the WISPR-O camera, which offers the best views of the trail. A comprehensive description of instrument specifications and capabilities can be found in \cite{Vourlidas15}.

\subsection{Ryabova Classical Cometary Model}   
\label{Sect: Ryabova model}

\subsubsection{Model Description}  
\label{sect:model_description}

The classical cometary model of the Geminid meteoroid stream (referred to as the ``Ryabova'' model in this manuscript) is the result of a near 40-years chain of publications, the last of which are \citep{2016MNRAS.456...78R, 2021MNRAS.507.4481R, 2022P&SS..21005378R}. Experiments with numerous and varied ejection models (collision, ejection in one point, cometary-like ejections, etc.) concluded that the cometary mechanism of the stream's formation fits best with the observed structure of the shower. A~cometary model of meteoroid stream formation has the following features: dust production increases toward perihelion; the ejection speed is smaller for larger particles; the ejections are mainly from the sunward side of the nucleus. 

For this study, we used the model calculated for the search for resonances in the Geminid stream \citep{2022P&SS..21005378R}, which in turn is the extension of \citep{2016MNRAS.456...78R}. The model consists of five model substreams with particles (spherical with density $\rho$ = 1~g~cm$^{-3}$) of fixed masses and corresponding spherical equivalent radii: $m_1 = 0.3$~g (4 mm), $m_2 = 2\times 10^{-2}$~g (2 mm), $m_3 = 3\times 10^{-3}$~g (900 $\mu$m), $m_4 = 3\times 10^{-4}$~g (400 $\mu$m) and $m_5 = 3\times 10^{-5}$~g (200 $\mu$m). Each of the fixed masses is a representative of the full mass interval. That is, $m_2$ represents interval [0.01, 0.1)~g, and $m_5$ represents [0.00001, 0.0001)~g. For this research, the model was expanded to include $m_6 = 3\times 10^{-6}$~g (90 $\mu$m) and $m_7 = 3\times 10^{-7}$~g (40 $\mu$m). We define $N_i$ and $N_{i+1}$ as the number of meteoroids in the neighboring mass intervals (say, $N_3$ and $N_4$ are for [0.001, 0.01) and [0.0001, 0.001)~g, with representing masses $m_3$ and $m_4$). We used 10,000 model meteoroids for each substream.

The ejection speed was calculated using the \citet{Whipple1951} formula, and the directions of the velocity vectors were distributed uniformly in the hemisphere facing the Sun. The equations of motion of meteoroids were integrated to the present epoch using the Everhart 19th-order procedure with variable step size. Planetary positions were taken from the JPL Planetary Development Ephemeris DE406 \citep{1998_Standish}. Gravitational perturbations from all planets, the Moon, and Pluto were taken into account as well as the {\it solar radiation pressure force} (SRP) and the {\it Poynting--Robertson} (PR) effect, including its corpuscular part. The starting epoch was taken $t_0$ = JD~1720165.2248 (perihelion passage), since the stream age was estimated in approximately 2000~yr \citep{1999SoSyR..33..224R}. 
The model sub-streams were generated during a single revolution of the parent body. Further details and references can be found in \cite{2016MNRAS.456...78R, 2022P&SS..21005378R}.

\subsubsection{Choice of the Reference Particle Mass}  
\label{sec:ryabova_ref_pcle}

We do not know the sizes or masses of the particles observed by WISPR, so we need to determine: should we involve the entire set of models we have, or is it enough to consider one or two dominating models? To answer this question, we can consider the action of radiation forces on meteoroids and the distribution of masses in the stream.

All integrations included the forces of SRP and PR drag. The radiative acceleration is proportional to $A/m$, where $A$ is the meteoroid cross-sectional area and $m$ is the mass \citep[][eq. (5)]{1979Icar...40....1B}. Generally, the smaller the particle, the larger the radiative force it experiences.

The SRP decreases the gravitational attraction of the Sun. The ratio of the SRP and gravitational forces has a commonly accepted designation $\beta$:
\begin{equation}
\beta = \frac{F_{r}}{F_{gr}} = 7.7 \times 10^{-5} \frac{A}{m}\;,
    \label{eq:beta}
\end{equation}
\noindent
where the numerical coefficient is valid for $A$ in [cm$^2$] and $m$ in [g]. The equation holds for particle radii above 1~$\mu$m. 
So, if we release particles from a parent body with zero ejection speed (a non-physical case), their semi-major axes will be larger than that of the parent body \citep[Table 1]{2022P&SS..21005378R}. At some $A/m$, SRP prevails gravitational attraction and the particle moves away from the Sun. This process was analyzed in detail by \citet{2021Icar..35413949M}, and for the Geminids the masses between $3.29 \times 10^{-16}$ and $1.18 \times 10^{-9}$~g (for density 1.45~g~cm$^{-3}$) turn out to be excluded from the stream completely. \citet[sect. 2.7]{2017P&SS..143..125R} estimated the minimal stream mass as $1.3 \times 10^{-13}$~g (for density 2.5~g~cm$^{-3}$). Smaller particles are swept out of the stream by SRP.

The PR force decreases both semi-major axis ($a$) and eccentricity ($e$) of a meteoroid orbit, causing the particle to spiral toward the Sun. This may not be the case for non-spherical particles with sizes up to tens of microns \citep{2010P&SS...58.1050K}, which are not considered in our models.
PR force has two components: photon and corpuscular. Elaborating the relations for PR-acceleration, \citet[p. 11]{1979Icar...40....1B} estimated the drag ratio of corpuscular to photon force to be of the order 0.1. Later \citet{2005dpps.conf..411R}, based on results of \citet{Mukai1982}, found that specifically for the Geminid stream the ratio is 0.4--0.7 (depending on material). Ryabova models here are calculated with a ratio of 0.4; the CS models (Section~\ref{Sect: CS-model}) do not take into account the corpuscular force. Detailed consideration, equations, and references can be found in \citep[sect. 3.1]{2022P&SS..21005378R}. 

The PR-effect counteracts SRP, and dominates it if given sufficient time. From \citep[Table 1]{2022P&SS..21005378R} it is easy to calculate that lifetime for $m_6$ and $m_7$-meteoroids\footnote {For short we will refer to `$m_4$-meteoroids' instead of `meteoroids with mass $m_4$' and similarly `$m_5$-stream' and so on.} is about 6 and 3~kyr respectively. For non-spherical particles $A/m$ increases by a factor of 1.5--1.75 \citep{2022P&SS..21005378R,2023VTGUM..83..151R}, decreasing lifetime in inverse proportion. Thus, we can conclude that the smallest masses in the stream are $m_5$ or $m_6$. In this manuscript we focus specifically on these masses, referring to them as the \textit{Ryabova $m_{5}$} and \textit{Ryabova $m_{6}$} models, respectively.

There is no consensus among researchers on the Geminid's density (see, for example, reviews in \citealt{1999SoSyR..33..224R}; \citealt[p. 101]{Jenniskens2024}). For our model, we adhere to the estimate $\rho$ = 1.0~g~cm$^{-3}$, but the parameter used for integration is $A/m$, and not $m$, thus the density is not important. 

The {\it differential mass distribution} in a meteor shower (or meteoroid stream) obeys a power law, generally accepted as:
\begin{equation}
f(m) = (s-1)m^{s-1}_0 m^{-s},    
\label{eq:f(m)}
\end{equation}
\noindent 
where $f(m)$ is the probability density function, $s$ is the differential mass index, and $m_0$ the limiting mass. From equation (\ref{eq:f(m)}) we can deduce \citep[sect.~2.3]{2021MNRAS.507.4481R} 
\begin{equation}
    N_{i+1}/N_i \approx 10^{s-1}.
\end{equation}

It is simple to show that the smaller the $m$, the larger the $N$. We observe brightness, so as a rough estimate of the relative measure of the trail `brightness' we use the total cross-sectional area of meteoroids $\Sigma A_i = A_i \times N_i$. Meteoroids in our model sub-streams do not interact, and can be removed from the stream only by the PR-effect. Thus, the initial and the final number of meteoroids in a sub-stream is constant.

Table~\ref{tab:total_cross-section} gives total cross-section areas for various $s$. These are relative values calculated for $N_3=1$. We must, however, still decide what value of $s$ we should use. We have numerous estimations of $s \approx 1.6-1.7$ for the {\it shower} observed at Earth for the peak of the shower. This is a very small sample of the Geminid {\it stream}, and its mass distribution may not be representative of the whole stream. 

After release from its parent body a meteoroid stream has an initial mass distribution of which our knowledge is still not very reliable \citep{Levasseur-Regourd2018}, and it can change greatly during evolution. Radiation forces disperse the stream, but they do not change $A/m$ of meteoroids. Collisions, though, can completely transform the mass distribution \citep{Jenniskens+2024}. Fortunately, this is not the case for the Geminids, due to their relatively small age \citep{1999SoSyR..33..224R}. 
Luckily, we need not consider these complicated issues, because \citet[tables 4--6]{2017P&SS..143..125R} compared the observed meteor shower with the model one, and estimated the modern Geminid stream mass index as $s = 2.2 - 2.4$, with the preference for the lower value. If so, the smallest particles ($m_5$ and $m_6$) dominate the trail brightness. 

\begin{table} [th]
\centering
 \caption{Total relative cross-section area (meteoroid density 1~g~cm$^{-3}$). Particle radii for each column are specified in the title}
 \label{tab:total_cross-section}
 \begin{tabular}{ccccccc}
  \hline
    s& $\Sigma A_1$ & $\Sigma A_2$ &$\Sigma A_3$ & $\Sigma A_4$ & $\Sigma A_5$ &$\Sigma A_6$ \\
     & 4~mm         &  2~mm        & 900~$\mu$m  & 400~$\mu$m   &   200~$\mu$m &   90~$\mu$m \\
\hline
1.5 &	0.054  &0.037  &0.025 &0.017  &0.012  &0.008 \\
1.6 &	0.034  &0.029  &0.025 &0.022  &0.019  &0.016 \\
1.7 &	0.022  &0.023  &0.025 &0.027  &0.029  &0.032 \\
1.8 &	0.014  &0.019  &0.025 &0.034  &0.046  &0.063 \\
1.9 &	0.009  &0.015  &0.025 &0.043  &0.074  &0.126 \\
2.0 &	0.005  &0.012  &0.025 &0.054  &0.116  &0.250 \\
2.1 &	0.003  &0.009  &0.025 &0.068  &0.185  &0.502 \\
\textbf{2.2} &	\textbf{0.002}  &\textbf{0.007}  &\textbf{0.025} &\textbf{0.086}  &\textbf{0.293}  &\textbf{1.001} \\
2.3 &	0.001  &0.006  &0.025 &0.108  &0.736  &1.998 \\
2.4 &	0.001  &0.005  &0.025 &0.136  &0.736  &3.986 \\
\hline
 \end{tabular}
\end{table}

The estimates described above were obtained with zero ejection speed. In the classical \cite{Whipple1951} formula we are using, the ejection speed depends linearly on $(A/m)^{0.5}$. It achieves 1.5~km~s$^{-1}$ for $m_5$-meteoroids and 2.2~km~s$^{-1}$ for $m_6$-meteoroids at the perihelion distance, so may not be ignored.

To understand how meteoroids of various masses become distributed in space under the influence of all factors, i.e. initial scheme of ejection, gravitational and radiational forces acting for 2000 yrs, we will consider the cross sections of our model stream in Section~\ref{sec:2.3.3_dist_in_space}. The consideration of these are crucial for understanding the positioning and distribution of dust observed in the WISPR trail.

\subsection{The Cukier-Szalay (CS) Model}  
\label{Sect: CS-model}

Unlike the Ryabova model, which was developed to analyze the Earth-observed Geminid shower features and based on the data of ground observations, the CS model is included here for comparison with PSP observations. Intentionally simplified, it presents three different schemes for the generation of the Geminids: `basic', `cometary', and `violent'. All three scenarios assume the Geminids were created in a single event (or single orbit) 2~kyr ago, and that any ongoing activity of Phaethon has not contributed to the stream. (The latter assumption seems particularly reasonable in light of recent studies such as \citealt{Zhang2023} and \citealt{Hui2023}.) In all models, the equations of the particle's motion were integrated considering gravitation forces for the Sun, Mercury, Venus, Earth, Mars, and Jupiter, as well as SRP and PR-effect (the latter did not include the corpuscular part). In the process, particles that were too close or distant from the Sun were removed. In all models, particle density was assumed to be $\rho$ = 3.205~g~cm$^{-3}$. All particles were weighted assuming $s = 1.68 \pm 0.04$ \citep{Blaauw2017}, but here we do not use the weight.
A brief description of each scenario is as follows, but we refer to \cite{Cukier2023} for a detailed breakdown of the assumptions and parameters of each model. 

In the {\it basic model} 10,000 particles were released from Phaethon at perihelion with zero ejection speed (an intentionally simplified case). Their $\beta$ values were uniformly distributed in [0, 0.052] interval. At the final moment, the number of particles was 9424.

In the {\it cometary model} a total of 100,000 particles were released (again with zero ejection speed), with 1,000 particles each at 100 discrete locations along the Phaethon orbit, distributed evenly in heliocentric distance and about perihelion therewith. In each location the maximal possible $\beta$ for elliptic orbits was found and the particle's $\beta$ values were uniformly distributed in $[0, \beta_{max}]$. Hereinafter we retain the author's name of the model, i.e. `cometary', although this model incorporates only one of the three features of a classic cometary model described in Section~\ref{sect:model_description}. Here, the final number of particles was 55,803.

The {\it violent model}, as the name suggests, simulates a violent creation mechanism for the Geminids in which all particles were ejected at a single instance at perihelion. Here, 100 discrete $\beta$ values were taken from 0 to 0.052 and 100 particles were simulated for each $\beta$. They were ejected isotropically with the speed obtained according to the statistical procedure described in \citep{1996Icar..120..212D}. In this case, the ejection speed depends on the particle mass, and its maximum is of the order 1~km~s$^{-1}$. In this model, 8,682 particles survived. We note here, as it becomes relevant later, that the Ryabova models also predict velocities in excess of 1~km~s$^{-1}$. However, these do so as part of a continual release process, with velocity decreasing with heliocentric distance. This is quite different to the instantaneous release mechanism of the CS Violent model.

In Table~\ref{tab:model-summary}, we provide a brief summary of these four models.

\begin{table}[ht!]
\centering
\caption{Summary of the model used for comparison with the WISPR trail.
\label{tab:model-summary}}
\begin{tabular}{|l|l|l|l|}
\hline
\textbf{Model} & \textbf{Meteoroid physical} & \textbf{Ejection scenario} & \textbf{Integration force model} \\
               & \textbf{properties}         &                            &                                  \\
\hline
\textbf{Ryabova classical}          & masses fixed:                  & Ejection around the orbit  & Gravitation from all planets,\\
\textbf{cometary}                   & $m_5 = 3 \times 10^{-5}$~g,    & from the sunlit hemisphere,& the Moon, and Pluto;         \\
$N = 10,000$ for           & and $m_6 = 3 \times 10^{-6}$~g,& ejection speed according   & radiation: SRP and PR -- photon \\
each mass                  & density = 1~g~cm$^{-3}$.       &  \citet{Whipple1951} formula.    & and corpuscular components. \\
\hline
\textbf{CS Basic }                  & $\beta$ in [0, 0.052],         & Released in perihelion &Gravitation from Mercury to \\
$N_{init} = 10,000$        & distributed uniformly,         & with zero speed.       &Jupiter, plus SRP and PR -- \\
$N_{final} = 9,424$        & density = 3.205~g~cm$^{-3}$.   &                        &photon component only. \\ 
\hline
\textbf{CS Cometary}                & $\beta$ in [0, $\beta_{max}$], & Released around the orbit,    &  The same as above. \\
$N_{init} = 100,000$       & uniformly; $\beta_{max}$ is    & 1000 particles in 100 points  &  \\
$N_{final} = 55,803$       & specific for each location,    & with zero speed.              &  \\
                           & density = 3.205~g~cm$^{-3}$.   &                               & \\
\hline
\textbf{CS Violent }                & $\beta$ in [0, 0.052],         & Ejection in perihelion,      & The same as above. \\
$N_{init} = 10,000$        & in 100 discrete values,        & isotropic, speed distributed & \\
$N_{final} = 8,682$        & density = 3.205~g~cm$^{-3}$.   & according \citep{1996Icar..120..212D}.      &\\
\hline
\end{tabular}
\end{table}

\subsection{Ground Video Observations}  
\label{sec:data_ground_obs}

We also selected two sets of ground video (Geminid meteor) observations to compare with WISPR observations, namely \citet{2003PAICz..91....1K} and \citet{2022A&A...667A.157B, 2022A&A...667A.158B}. Both catalogs are freely available. 

The first catalog \citep{2003PAICz..91....1K} contains data on 71 Geminid meteors obtained within the double-station video observation in 1999 and 2001 at the maximum (December 13--14) of the shower. The photometrical mass of meteoroids was calculated by integration of the meteor light curve and was obtained within $[0.00068, 0.10]$~g. The maximum brightness of the meteors adjusted to the distance of 100~km is within $[-1.4, +4.3]$ magnitude. \citet{2017P&SS..143...89H} analyzed Geminid orbits from the six known video catalogs. The Czech database of video meteors \citep{2003PAICz..91....1K} was found to be the only one from this set that does not underestimate the determined velocity. The high quality and precision of this data sample is achieved through a rigorous processing approach: it contains only good quality records, and raw data were measured manually. 

The second catalog \citet{2022A&A...667A.157B, 2022A&A...667A.158B} contains data on 38 Geminid fireballs observed by the European Fireball Network in 2017--2018 (December 11--14). The observations were made by digital photographic cameras. Here again every meteor was processed manually (partly semi-automatically) aiming at the best possible accuracy of the data. The maximum brightness of the meteors is in the range $[-11.5, -3.1]$ of the absolute (100~km distance) magnitude. The photometrical mass of the meteoroids is within [0.17, 1600]~g. The authors investigated the orbital configurations of the Geminids meteors\footnote{Technically speaking, meteors are the ``the light and associated physical phenomena'' \citep{2016JIMO...44...31B} resulting from a meteoroid, and do not as such have, e.g. mass or size; nevertheless, for convenience we refer to shower meteoroids as ``meteors''.} and their inter-dependencies \citep[4.1.1] {2022A&A...667A.158B} and made a comparison with Ryabova's model \citep{Borovicka2023}. This study noted an unusual feature of this sample: the expectation is that larger meteors should have larger semi-major axes and eccentricities because of the PR-effect. However, contrary to expectation, exactly the opposite effect was observed. \citet{1993mtpb.conf..193S} has already noted this unusual trend by analyzing data from photographic fireballs. The reason behind this is still unclear.

\section{Methodology}

\subsection{Reference System for Visual Analyses of WISPR FOV}
\label{sec:method-visual-analyses}
Much of this work relies on simulations of Geminid models in the WISPR FOV, which are used for two purposes. First, we use them to compare the models visually with the appearance of the WISPR trail. Second, we use them to obtain numerical values for trail properties -- specifically, the trail (model and observations) width, and apparent offset from the orbit of Phaethon. These simulations were produced only at certain `snapshot' time instances in each Encounter, with details to follow. 

All Geminid models discussed here provided a series of orbital parameters that describe the orbits of individual particles or groups of particles. With such data, we can use World Coordinate System \citep[WCS,][]{Thompson2006} PSP/WISPR metadata to plot the apparent trajectories of these particles across a simulated (or actual) WISPR FOV to generate full-sequence simulations for all WISPR observations. 

Due to similarities in the six PSP orbit configurations, described in Section~\ref{sec:data}, simulation sequences were produced for only one Encounter per orbit configuration. For additional brevity, we consider only simulations from Encounters 4, 6, and 13, which show the largest orbit-to-orbit visual differences in WISPR's view of the models. 

\begin{figure}[ht!]
\centering
\includegraphics[width=0.98\textwidth]{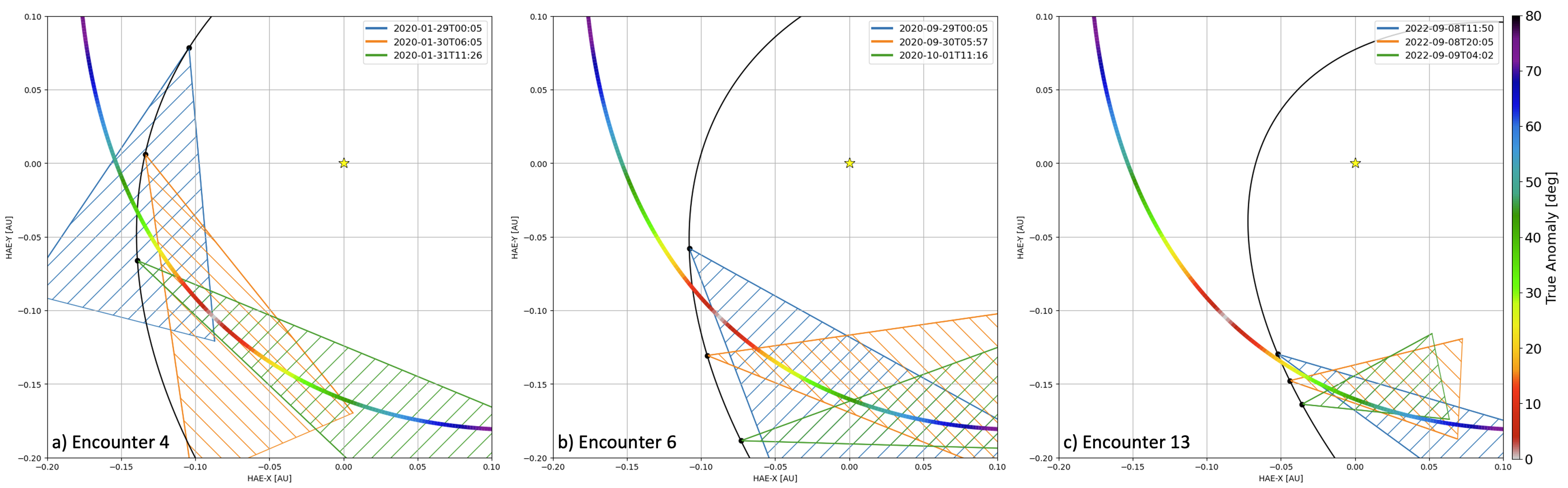}

\caption{Orbital positions and effective FOV for the simulation images shown in later Figure~\ref{fig:ryabova_m5_m6_3x3} and other similar figures, for Encounters (a) E4; (b) E6; and (c) E13. The solid black line represents the orbit of PSP relative to the Sun (yellow star), on a Heliocentric Aries Ecliptic (HAE) x-y grid (units of au), as viewed from above the ecliptic plane. The color bar follows the orbit of Phaethon, with the color encoding representing TA (degrees). The shaded triangles represent the effective FOV of WISPR at: (a) 2020-01-29 00UT (blue), 2020-01-30 06UT (orange), and 2020-01-31 11UT (green); (b) 2020-09-29 00UT (blue), 2020-09-30 06UT (orange), and 2020-10-01 11UT (green); (c) 2022-09-08 12UT (blue), 2020-09-08 20UT (orange), and 2022-09-09 04UT (green).}
\label{fig:orbit_sims}
\end{figure}

Later sections will present Figures comprising a selection of frames from the full imaging sequence simulations of every model/scenario within E4, E6, and E13, with a common layout for each Figure. Figure~\ref{fig:orbit_sims} presents a 2D orbital overview, and effective FOVs of the WISPR-O detector on the selected dates and times, corresponding to three positions along the PSP orbit in E4 (Panel a), E6 (Panel b) and E13 (Panel c). These dates/times are those for which we will present simulation views. The PSP orbit is shown in black, and Phaethon's orbit is shown by a color bar representing TA. The yellow star indicates the Sun, and the entire view is from the perspective of an observed situated above the ecliptic plane. Thus, for each Encounter in consideration (E4, E6, E13) we will present three snapshots from the simulation for that Encounter, for a total of nine panels per Figure. 

There is no particular rationale for selecting the specific date/times that we have chosen. The passage of the Phaethon trail through the WISPR-O FOV encompasses approximately 90 individual WISPR frames over an approximately two-day time period. The chosen frames were selected semi-randomly, with the only criteria being that a substantial portion of the trail/simulation was in the FOV at that time, and that they were sufficiently spaced temporally to show evolution of the viewing geometry. 

These simulations are physics-free in the sense that no adjustments are made for particle size, density, phase angle effects, or any other physical property that may affect particle visibility. These parameters are beyond the scope of this investigation. The simulations are simply created by recursively tracing the path of each model particle across that given FOV at that selected date/time. The intensity of any given pixel in each simulation is directly proportional \textit{only} to the number of particles that have crossed over that particular pixel (effectively a line-of-sight density). In later analyses, a uniform random synthetic noise floor is added to the simulations to remove most of the faint orbit tracks and leave only the most visually dense portion of the models (cores). Further details of this are provided by Appendix~\ref{app:A}.

\subsection{Trail Measurements}
\label{sec:methods-trends}
In Section~\ref{sec:results-trends} we explore the numerical trends in both the observed and model trail properties as functions of TA along Phaethon's orbit. Specifically, we look at the apparent (i.e., in the FOV) offset of the trail from Phaethon's current orbit, and the apparent observed width of the trail, both as functions of TA. For the latter in particular, we will frequently comment that the apparent trail width is a function of the \textit{noise level} of the WISPR observations (both the actual and simulated observations). That is, we have no way of knowing exactly how wide the trail is, or even if ``width'' is an appropriate metric for something that is likely part of a continuous distribution. We can only observe and analyze whatever is above the noise threshold of the instrument. Thus, the purpose of these analyses is to show only that \textit{trends} observed by WISPR are consistent with those seen in the models from a similar PSP-based viewpoint. 

The methodology for determining these quantities is effectively the same as employed in \cite{Battams2022}. In short, the data are stacked to increase the trail signal, and then smoothed to decrease noise, thus enabling the extraction of cross-trail profiles. These profiles can then be fitted with Gaussians in order to estimate the width (full-width half-maximum), and the distance between Phaethon's orbit and the Gaussian peak (the `offset'). The only substantial change was to sample the trail at a much higher angular resolution in TA than explored by \cite{Battams2022}. This technique was applied to both the WISPR data and the model simulations in the FOV, after first adding a random noise floor to better mimic the properties of the WISPR data, and remove faint particle tracks. This approach enabled estimates of the widths of the densest portion of the trail, as well as the distance between Phaethon's orbit and the trail (core). Uncertainties are based on all observations at a given TA across all time instances, and are driven primarily by the highly variable viewing geometries across each encounter. Other details pertaining to the preparation of the model data are provided in Appendix~\ref{app:A}.

\subsection{Phaethon-orbit Centered Coordinate System}
\label{sec:methods-coord-scheme}
Interpretations of these offset results also require an understanding of the dust distributions in the various models. In Section~\ref{sec:results} we present cross-sections of the models to help explain the observed properties of the WISPR trail, and the Earth-encountering Geminids. These cross-sections are depicted in terms of a $\xi$--$\eta$ system, which we define as follows. 

\begin{figure}[th]  
  \begin{center}
     \includegraphics[width=50 mm]{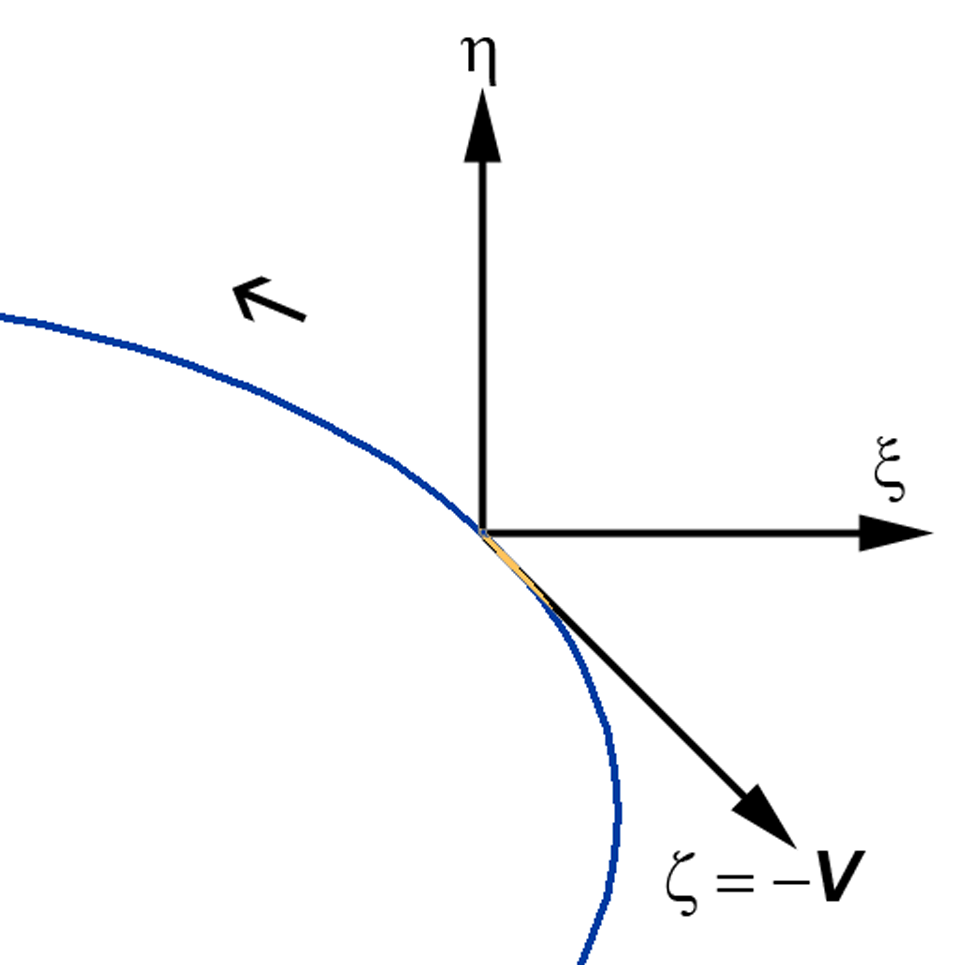}
\caption{
Phaethon-orbit centered coordinate system $\xi, \eta, \zeta$. 
}
    \label{fig:0}
  \end{center}
\end{figure}

Let the reference orbit be given by its velocity $\mathbf{V}$ and position $\mathbf{r}$ vectors at a given TA, $\upsilon$ (degrees). The vectors are given in standard heliocentric ecliptic system. Let the origin of a rectangular right-handed coordinate system lie at this point. The $\xi$-axis (abscissa) lies in the plane of the reference orbit and is directed away from the Sun. The $\zeta$-axis\footnote{The $\zeta$ coordinate is not used in this publication, but is included here for completeness. The definition of the positive/negative direction is a historical carryover that has no bearing on this study.} (applicate) is along $\mathbf{-V}$, and the $\eta$-axis (ordinate) completes the system. This is a modification of an object-centric system, but we use it as a Phaethon-orbit-centered system, illustrated by Figure~\ref{fig:0}. We can then place it on any desired TA. The cross-section of the model stream, in a plane perpendicular to the Phaethon orbit's velocity vector, where $\xi, \eta$ are the model meteoroid orbit's node coordinates, and the Phaethon orbit pierces the plane $\xi, \eta$ at (0, 0), allows us to see density distribution of orbits across the stream. The nodes with $\eta > 0$ are located above Phaethon's orbital plane and vice versa. The nodes with $\xi >0$ lie outside the Phaethon orbit, and those with $\xi<0$ inside it. 

\section{Results} \label{sec:results}

We now present the results of our investigation, beginning with a very brief update to \cite{Battams2022}. This is followed by an exploration of the cross sections of the model orbits to explore their distribution in space. We will then present the model simulations in the WISPR FOV, and finally our results looking at trends in model and trail properties as a function of TA.

\subsection{PSP Latest Trail Observations / Update to Previous Work}\label{sec:analysis-psp-update}  

First, we provide here a brief incremental update on the WISPR observations of the trail, following the identical procedures as described in previous publications. The latter of those, \cite{Battams2022}, incorporated observations through E9 (2021 Aug 04 -- 15), estimating the mass of the trail to be $\sim10^{10}$~--~$10^{12}$~kg with visual magnitude V$\sim$16.1$\pm$0.3 per pixel, corresponding to a surface brightness of 26.1~mag~arcsec$^{-2}$. In this manuscript, we expanded our dataset to include E10 -- E16 (through 2023 June 16 -- 27). As illustrated in Table~\ref{tab:enc-overview}, despite this data set encompassing six encounters, this only represents one new orbital regime for the spacecraft. 

In this regime, PSP is within 0.0136~au of the Phaethon orbit -- approximately half the distance from the previous orbit regime (E8--E9, 0.0277 au). Accordingly, the view of the trail evolved once again. Most notably it appears substantially wider and fainter, as the same level of signal is now spread over many more pixels, reducing the visibility of the trail. Even stacking of the data results in a very noisy signal that is challenging to fit reliably. This is unfortunate but not unexpected; in the most extreme hypothetical case, if PSP were to fly directly into the dust trail, WISPR would certainly lose sight of it as the signal would encompass the entire FOV. Thus, while WISPR observations beyond E16 are available, they are not discussed here for these reasons. 

Attempts were made to perform photometric updates to \cite{Battams2020, Battams2022} using the same technique as those studies. However, the uncertainties in returned values were so large as to not provide any meaningful updates to those already published. Thus, we do not offer any further refinements to the trail surface brightness or mass estimates in this manuscript. 

We next present an exploration of the distribution in space of the various Geminid models under consideration, employing the coordinate scheme defined in Section~\ref{sec:methods-coord-scheme}. Such explorations of the distribution of orbits in the models are crucial for the later interpretation of both the visual features and trends we observe in the WISPR data and models. Following these distribution results, we then present the WISPR FOV simulations for each model.

\subsection{Ryabova Models: Distribution in Space} 
\label{sec:2.3.3_dist_in_space}

\begin{figure}[th]  
  \begin{center}
     \includegraphics[width=70 mm]{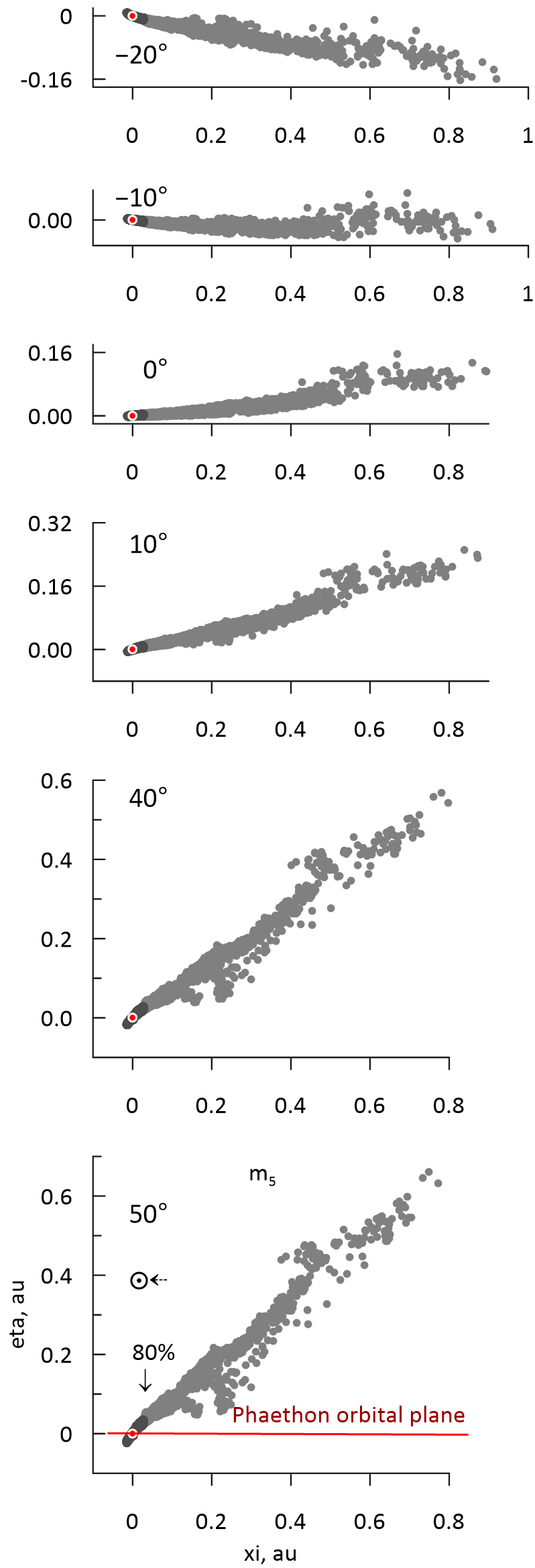}
\caption{
Cross-sections of a model Geminid $m_5$-sub-stream consisting of 10~000 of orbits at various places of the reference orbit. Their nodes are shown by gray dots. The densest parts containing 80\% of nodes are shown by dark gray dots. The TA $\upsilon$ in degrees (from $-20^\circ$ to $50^\circ$) is given on the each plot. The Phaethon orbit is located at point (0,0), shown by the large red dot. The plots are given on the same scale. The axes are as illustrated in Figure~\ref{fig:0}.
}
    \label{fig:1}
  \end{center}
\end{figure}

Figure~\ref{fig:1} shows cross-sections for an $m_5$-sub-stream at the TA values indicated on each panel. We see that the 80\%-`core' of the stream (i.e. 80\% of the stream meteoroids) is concentrated in the vicinity of the parent body orbit, while the 20\%-`tail' extends for almost 1~au. The limit of 80\% was chosen rather arbitrarily via trial and error. Moreover, for simplicity, we just cut the plot at the 80\%-level for $\xi$. 

\begin{figure}[ht]  
  \begin{center}
     \includegraphics[width=140 mm]{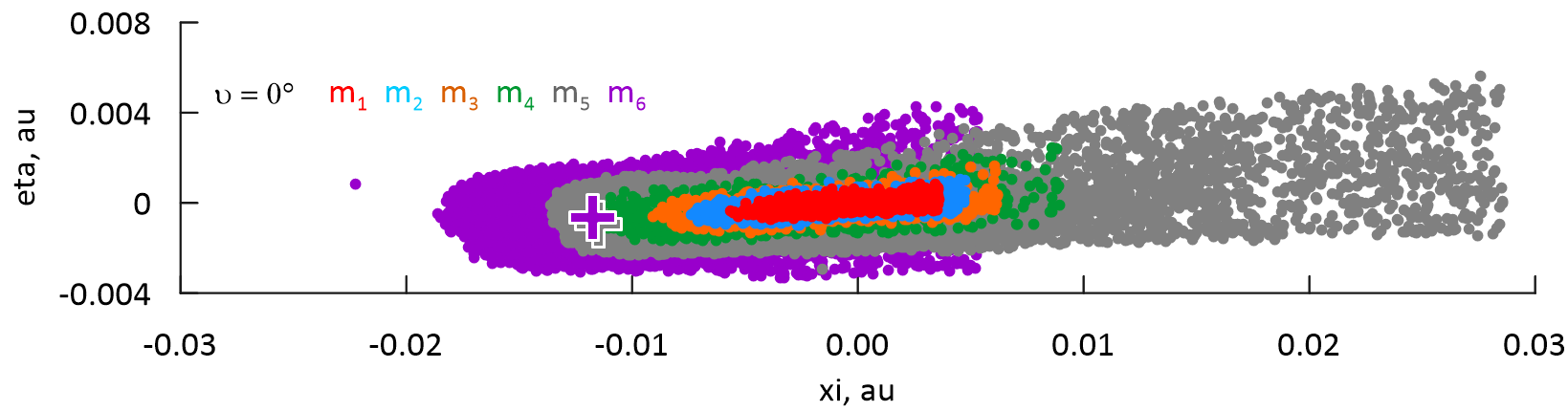}
\caption{
Cross-sections of a model Geminid sub-streams of various meteoroid masses at perihelion of the reference orbit. Here only 80\%-cores are shown. The large crosses designate the density maxima.
}
    \label{fig:2}
  \end{center}
\end{figure}

\begin{figure}[ht]  
  \begin{center}
    \includegraphics[width=175 mm]{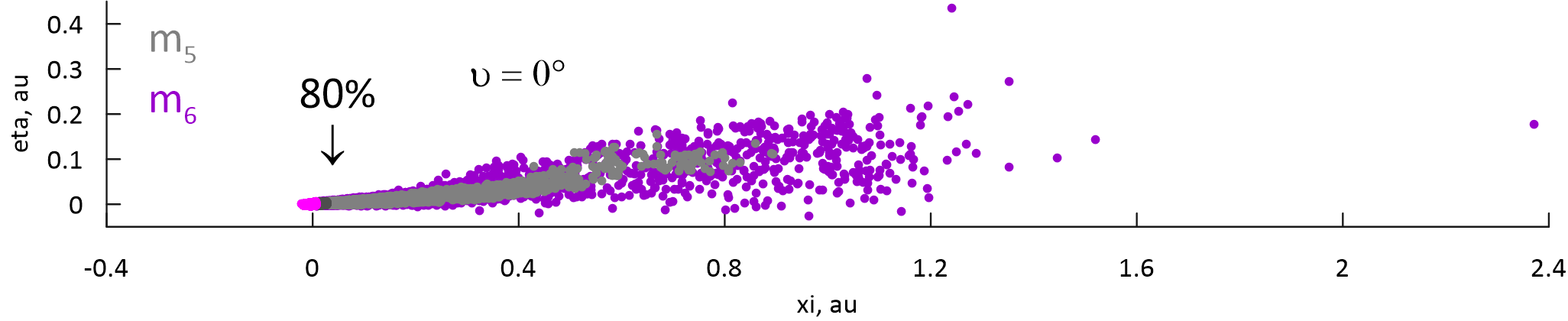}
\caption{
Cross-sections of the whole model Geminid $m_5$- and $m_6$-streams at $\upsilon = 0^\circ$. Their 80\%-cores are also shown by dark gray ($m_5$) and light-purple ($m_6$) colors.
}
    \label{fig:3}
  \end{center}
\end{figure}

It can be predicted without modeling that cross-sections for sub-streams $m_1$ -- $m_4$ must be inside of the $m_5$-stream due to their lower ejection speed. This is demonstrated in Figure~\ref{fig:2}. However, the $m_6$ cross-section does appear different, but only if we consider 80\%-cores. For the whole $m_5$- and $m_6$-streams, the full picture is as expected (Figure~\ref{fig:3}). This is explained by the orbital contraption due to PR-effect overcoming the initial dispersion for this part of the orbital space of $m_6$-meteoroids.

\begin{figure}[ht]  
  \begin{center}
     \includegraphics[width=152 mm]{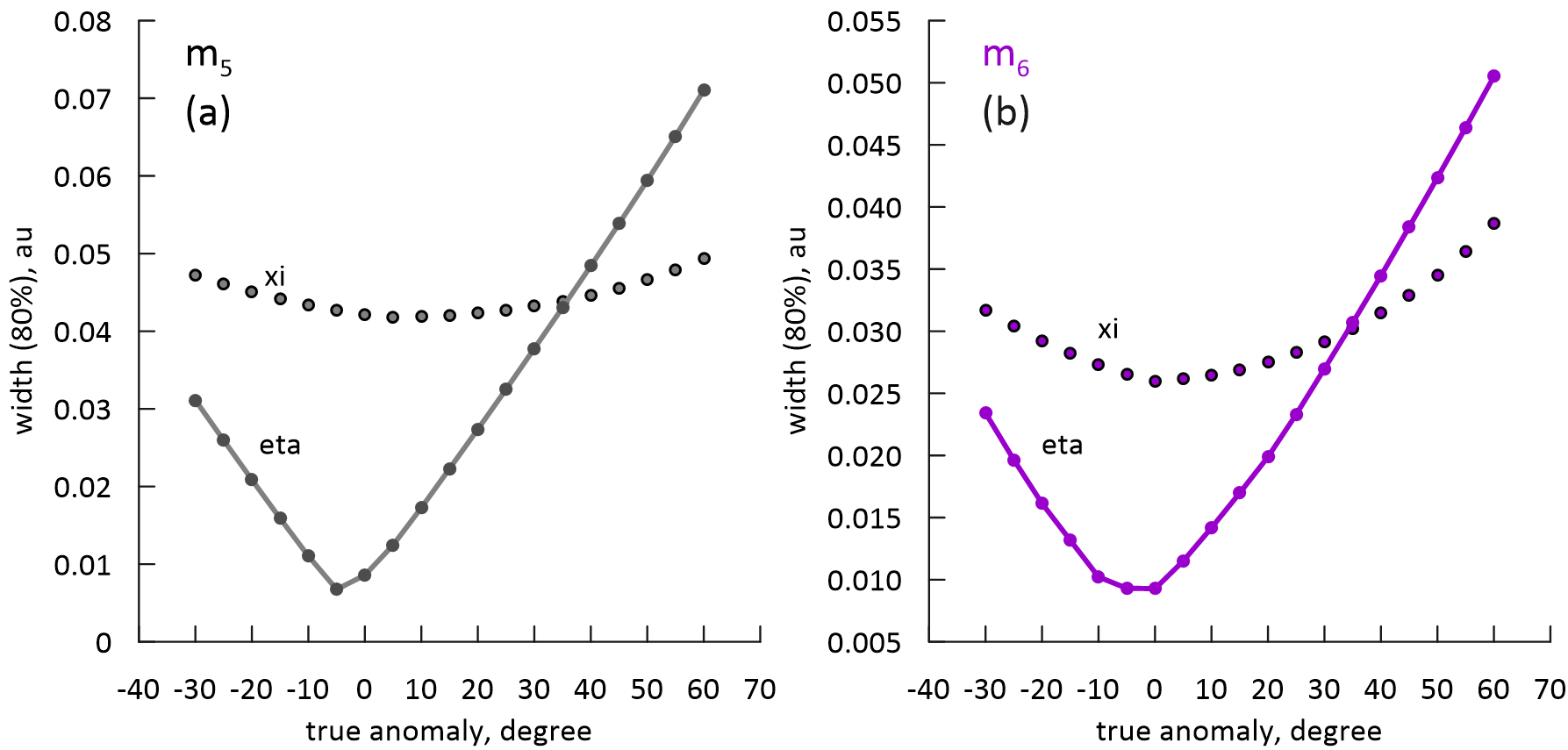}
\caption{Width of the 80\%-core for model Geminid sub-streams (a) $m_5$ and (b) $m_6$. The width was measured along $\xi$ and $\eta$ axes of the streams' cross-sections.
}
    \label{fig:4}
  \end{center}
\end{figure}

The width of the 80\%-core depends on the TA (Figure~\ref{fig:4}), being the smallest near perihelion and largest near aphelion. This is expected for a bunch of high-eccentric orbits with limited dispersion in $\Delta a$ and $\Delta e$ and follows from the formulae for perihelion distance $a(1-e)$ and aphelion distance $a(1+e)$.

\begin{figure}[ht]  
  \begin{center}
     \includegraphics[width=152 mm]{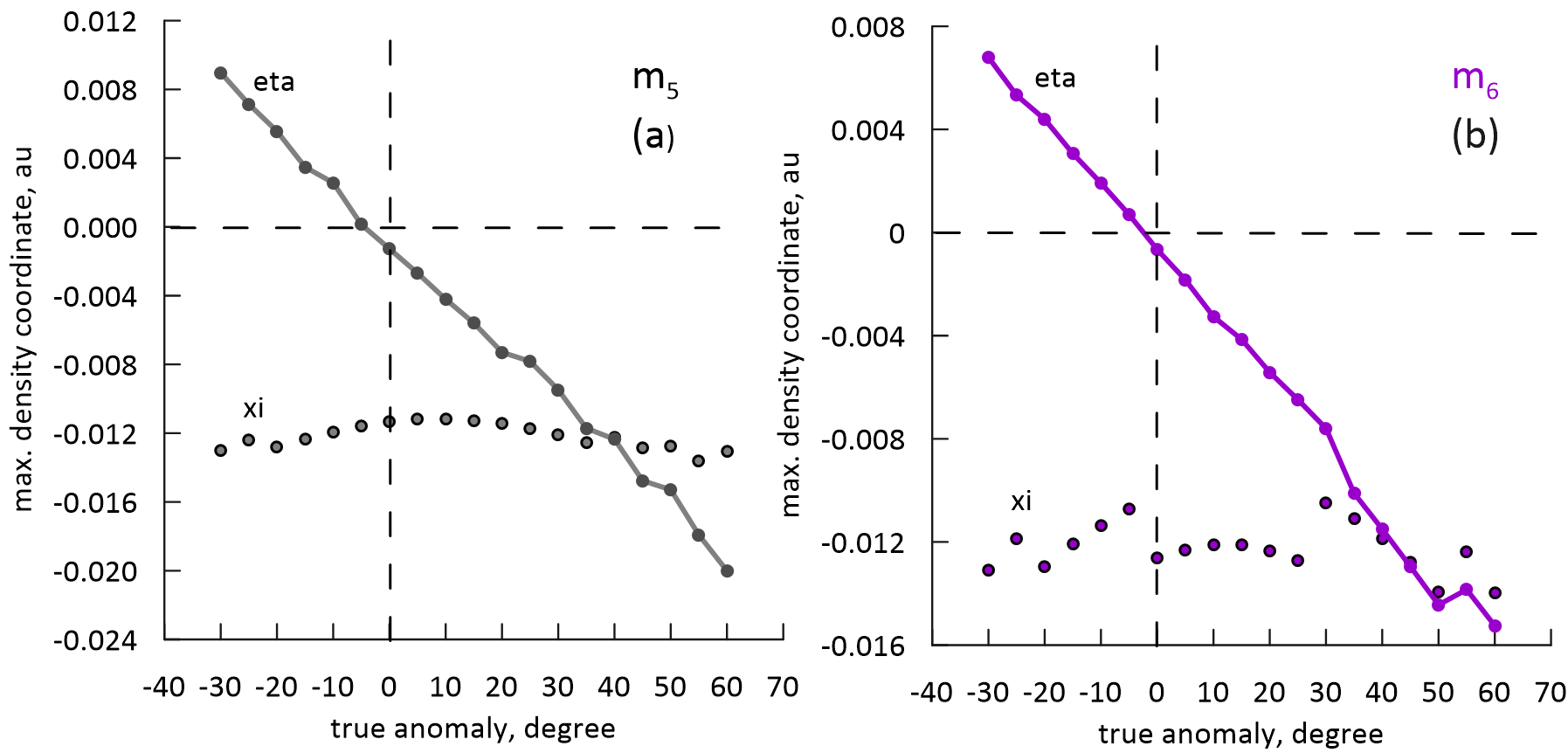}
\caption{Position of the maximum density in the cross-sections of model Geminid sub-streams (a) $m_5$ and (b) $m_6$. Let us remind you: $\xi$ shows the position inside (minus) or outside (plus) the Phaethon orbit, while $\eta$ shows whether the position is above (plus) or below (minus) the Phaethon orbital plane.
}
    \label{fig:5}
  \end{center}
\end{figure}

The three dimensional (3D) shape of the Geminids is rather complicated, resembling a sea shell \citep{2007pimo.conf...71B}. The most concentrated filament passes close to and inside Phaethon's orbit (Figures~\ref{fig:1}, \ref{fig:2}). Before perihelion it passes above the Phaethon orbital plane, near perihelion it dives under the plane (Figure~\ref{fig:5}), and near aphelion again goes up. So in perihelion the stream twists like a M\"{o}bius strip. The jumps in the $\xi$-curve on the Figure~\ref{fig:5}b are explained by this. They are also present in Figure~\ref{fig:5}a, but are very weakly expressed. The shape of the stream is determined by the initial distribution of particles and the subsequent evolution of their orbits \citep[in detail see in][2.1]{2022P&SS..21005378R}. The mentioned 20\%-tail consists of meteoroids with the high-$i$ and high-$a$ (inclination $i$ depends on $a$) orbits. The precession of orbits (i.e. circulation of the argument of perihelion) depends on ($a, e, i$) and occurs at different speeds, resulting in a special 3D-shape of the model stream. The minimum in the model stream width (Figure~\ref{fig:4}) occurs several degrees in TA of the Phaethon orbit before perihelion, precisely due to precession.

Returning to Figure~\ref{fig:5}, we re-emphasize that the densest part of the model stream is always inside the Phaethon, or more generally the \textit{parent}, orbit. Before perihelion it is located above its orbital plane, but after perihelion it is below the plane. This observation is important to keep in mind when we later discuss trends in the positioning (offset) of the WISPR trail relative to Phaethon's orbit.

\subsection{CS Models: Distribution in Space} 
\label{sec:cs_distribution_in_space}

\begin{figure}[ht]  
  \begin{center}
     \includegraphics[width=0.9\textwidth]{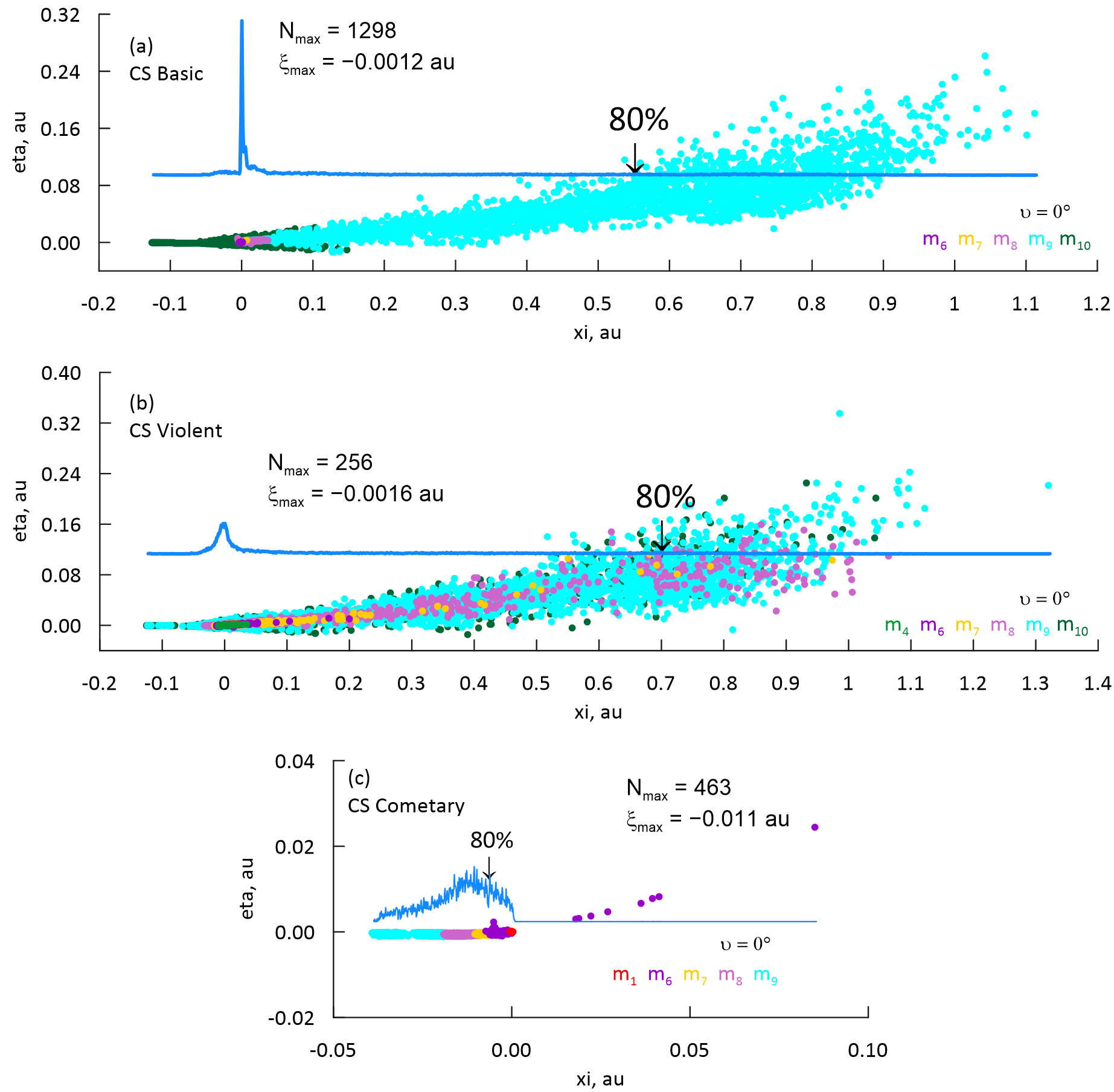}
\caption{Cross-sections of (a) the CS Basic; (b) the CS Violent, and (c) the CS Cometary models. The nodes of model meteoroid orbits of various masses ($m_1, \ldots, m_{10}$) are shown by dots of the corresponding color. The solid blue line presents the nodes density distribution along $\xi$-axis. Location of the peak $\xi_{max}$ relative to the Phaethon node (0,0) and the peak number of meteoroids ($N_{max}$) are provided on each panel. The 80\%-arrow indicates the border between the dense part of the stream containing 80\% of particles and the rarefied tail. Note that the scale in $\xi$ and $\eta$ axes on (c) is not the same unlike (a) and (b).}

    \label{fig:6}    
  \end{center}
\end{figure}

We can consider the CS models in the same manner as the Ryabova model. In Figure~\ref{fig:6} we show cross sections of the CS models at perihelion, for various masses. The mass designations on these panels are the same (including the color code) as for the Ryabova model, i.e. $m_1$ for [0.1, 1.0)~g, \ldots , $m_{10}$ for [$0.1\times 10^{-10}, 0.1\times 10^{-9})$~g. The Ryabova and CS models differ in their representation of the mass interval (mean mass versus continuous distribution), but the difference is not significant in this case. Also, we note that CS models assume a particle density of $\rho = 3.205$~g~cm$^{-3}$, as opposed to the $\rho = 1.0$~g~cm$^{-3}$ in the Ryabova model. These higher density particles therefore have a smaller cross-sectional area, so the radiation forces for them are smaller. Furthermore, the PR effect for the Ryabova model is stronger than the CS models by 40\%. This roughly corresponds to an increase in $A/m$ by 1.4 times. Thus, the Ryabova $m_5 = 3\times 10^{-5}$~g is dynamically equivalent to a CS mass of $1.1\times 10^{-6}$~g. 

This Figure~\ref{fig:6} shows us that the peak density is located inside the Phaethon orbit for all models. The CS Basic model demonstrates what happens with meteoroids ejected at perihelion with very small (here zero) speed: for the small meteoroids, the perceptible mass separation takes place, while the large meteoroids remain concentrated in the peak inside the parent orbit. Practically all meteoroids of the CS Cometary model relocated inside the orbit of Phaethon. As discussed later, this model therefore does not perform well at reproducing Earth-encountering meteors -- indeed, \cite{Cukier2023} state that none of their models created Earth-impacting meteoroids -- but the CS Cometary model in particular was unable to place any particles close to Earth. The CS Violent model has greater dispersion and less mass separation, as expected for such a scenario.

\subsection{Ground Video Observations: Distribution in Space}
\label{sec:res_ground_dist_in_space}
\begin{figure}[ht]  
  \begin{center}
     \includegraphics[width=0.5\textwidth]{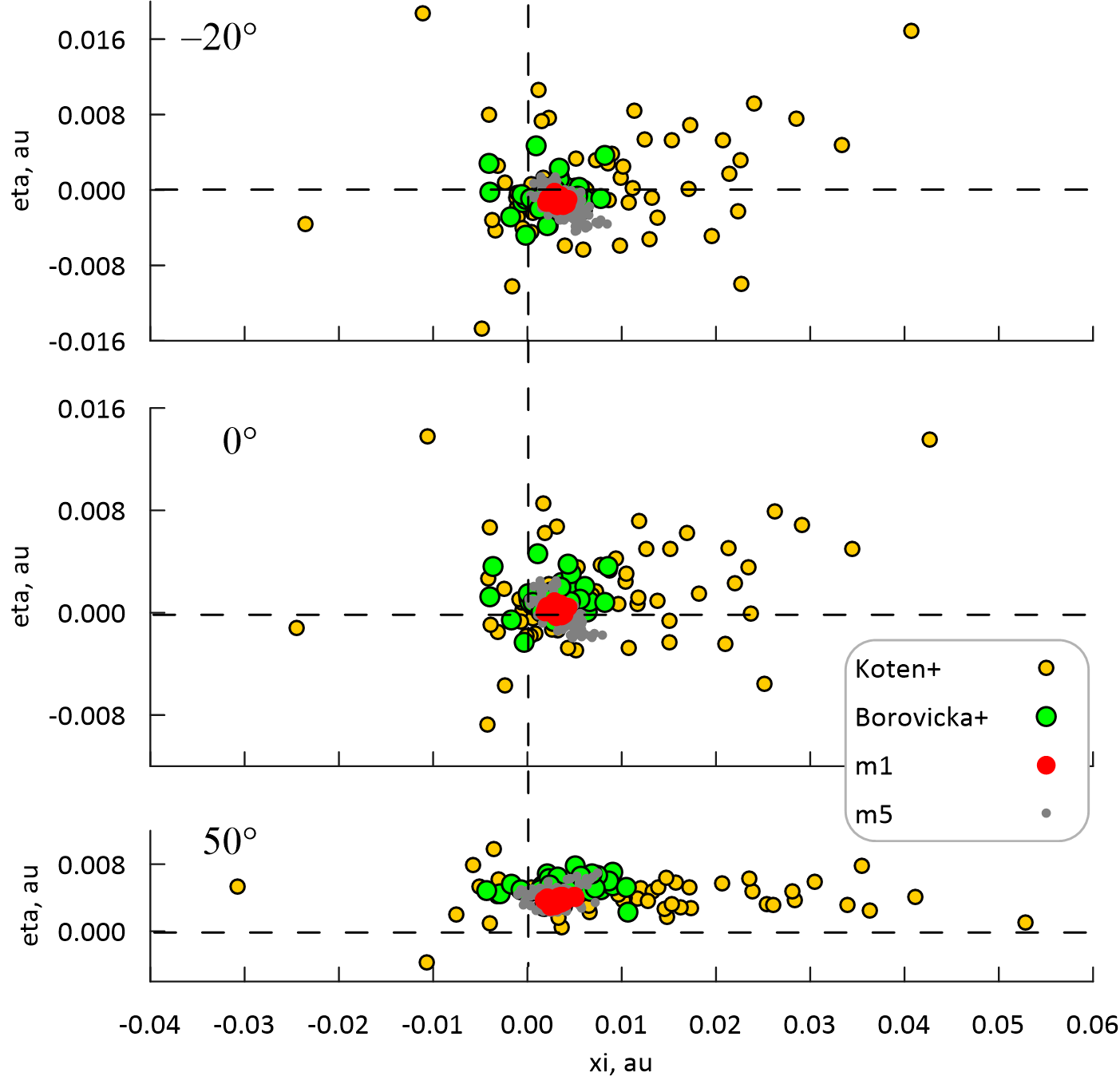}
\caption{
Cross-sections of the two samples of observed meteor orbits and two samples of Ryabova model meteor shower orbits (40 orbits with masses $m_1$ and 40 orbits with mass $m_5$) in three true anomalies of the reference (Phaethon) orbit: $-20^\circ$ (top), $0^\circ$ (middle), $50^\circ$ (bottom). The coordinate system is centered on the Phaethon orbit and is described in Sect.~\ref{sec:methods-coord-scheme}. 
}
    \label{fig:7}
  \end{center}
\end{figure}

Finally we also consider the distribution of the ground-observed meteors. It is important to note that observed meteors are only a tiny part of a meteoroid stream captured by the Earth piercing the stream, leading to a strong observational bias in those observations. Therefore, observed meteors should be compared with the model {\it shower} meteors, i.e. the model stream meteoroids registered at the Earth. Figure~\ref{fig:7} shows the location of the observed and Ryabova model meteor shower orbits in the cross-section centered on the Phaethon orbit. We consider the meteoroids having nodes on the distance 0.005~au within the Earth's orbit as Earth-intersecting, i.e. as model shower meteors. For the considered 10000-particle models we found 205 $m_1$-meteors and 43 $m_5$-meteors. All samples, both observed and model, show the same pattern of location: they are (mostly) outside the Phaethon orbit; they are (mostly) below the Phaethon's orbital plane before perihelion and obviously higher than it after perihelion. The dispersion of the model orbits is smaller as expected, but in general the agreement is good. 

\begin{figure}[ht]  
  \begin{center}
     \includegraphics[width=0.6\textwidth]{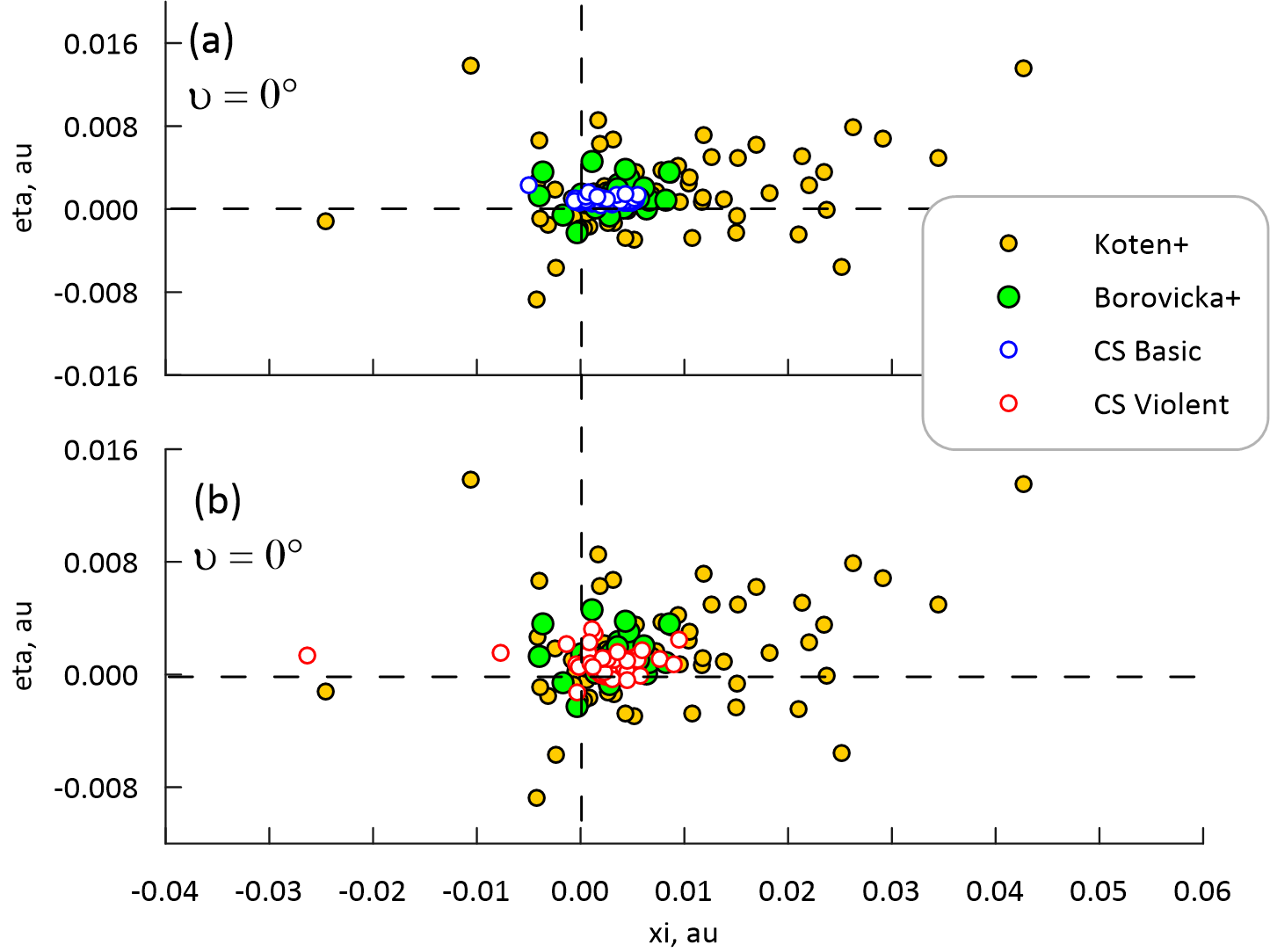}  
\caption{Cross-sections of the two samples of observed meteor orbits and two samples of CS model meteor shower orbits at $\upsilon = 0^\circ$. (a) CS Basic, 70 meteors, and (b) CS Violent, 52 meteors.} 
    \label{fig:8} 
  \end{center}
\end{figure}

The same is observed for the CS models, as shown in Figure~\ref{fig:8}, for both CS Basic and CS Violent models. The CS Cometary model does not contain Earth-encountering meteoroids. Figure~\ref{fig:8} presents plots only for $\upsilon = 0^\circ$ for brevity, because the plots for $\upsilon = -20^\circ$ and $\upsilon = +50^\circ$ are similar to those in Figure~\ref{fig:7}. This is expected, since the final orbital space of the Ryabova model contains almost the entire orbital space of the CS model. So the meteoroids encountering the Earth -- and this is a very strong condition -- have more or less similar orbits.

\begin{figure}[ht]  
  \begin{center}
     \includegraphics[width=0.5\textwidth]{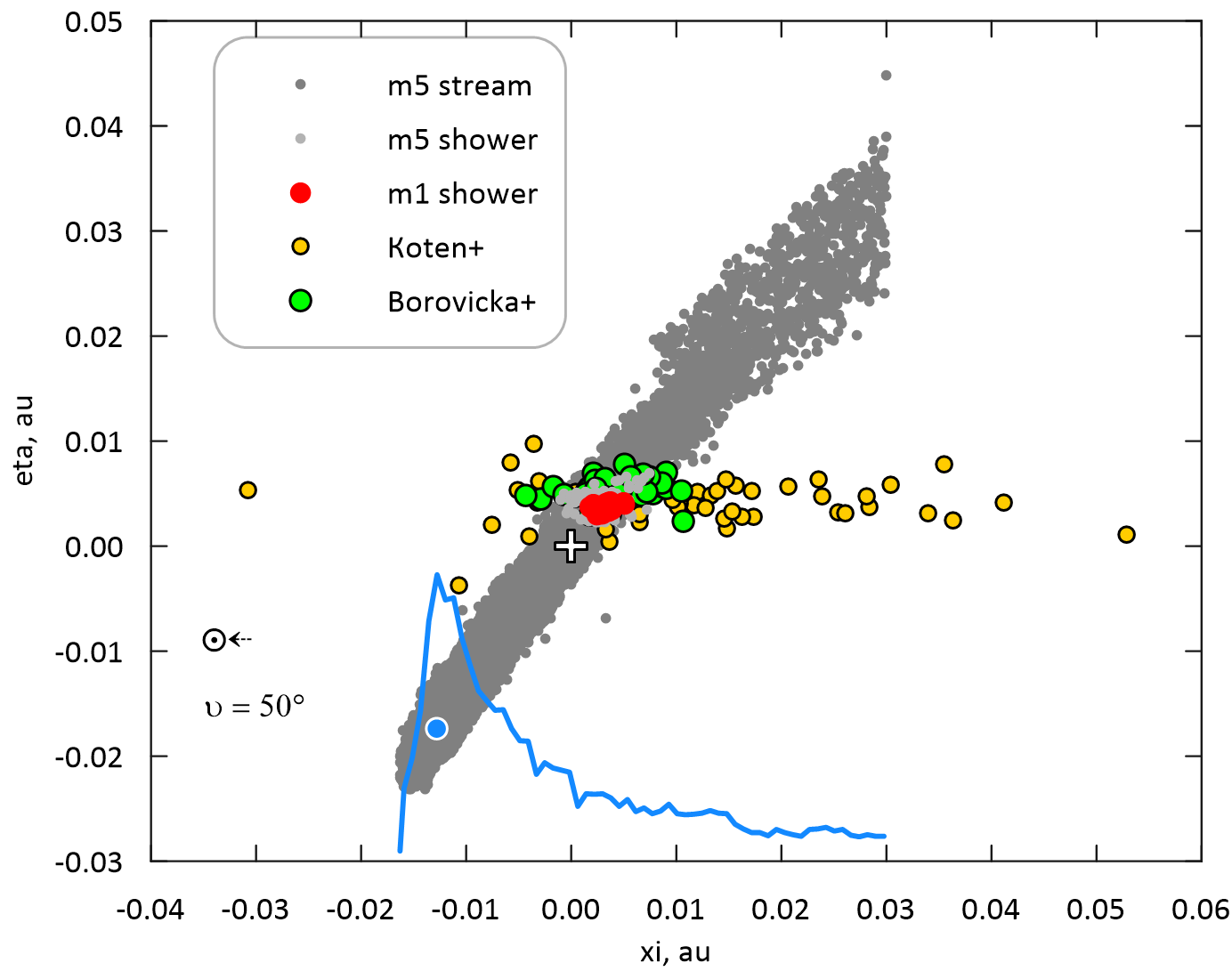}
\caption{
Cross-sections of the two samples of observed orbits, two samples of model {\it shower} orbits (the same as in Figure~\ref{fig:7}) and Ryabova model stream orbits (80\% of meteoroids of $m_5$-sub-stream) for the TA $\upsilon = 50^\circ$ of the reference (i.e. Phaethon) orbit. Coordinate system is described in Sect.~\ref{sec:methods-coord-scheme}. The large cross designates the Phaethon's node. The blue circle marks the place of the maximal density in the stream and the blue curve shows the particle nodes density distribution in the model stream along the $\xi$-axis. The direction to the Sun is shown on the plot.
}
    \label{fig:9}
  \end{center}
\end{figure}

To demonstrate the impact of meteoroid (shower) observational bias, we refer to Figure~\ref{fig:9}. We see that the most dense filament of the Ryabova model stream is located far below Phaethon's orbit and closer to the Sun, and the shower orbits (both model and observed) are much farther from the Sun and above the reference orbit. Thus, it is not possible to truly assess the entire population of a meteoroid stream based solely on observations of its shower meteoroids. 

\subsection{WISPR Simulations of Models}\label{sec:qualitative_comp}

The following sections present visualization of the various Geminid simulations in the WISPR FOV at specific time instances in each PSP Encounter, as described in Section~\ref{sec:method-visual-analyses}.

\subsubsection{Ryabova Model Simulations}\label{sec:ryabova_sims}

\begin{figure}[ht!]
\centering
\includegraphics[width=0.99\textwidth]{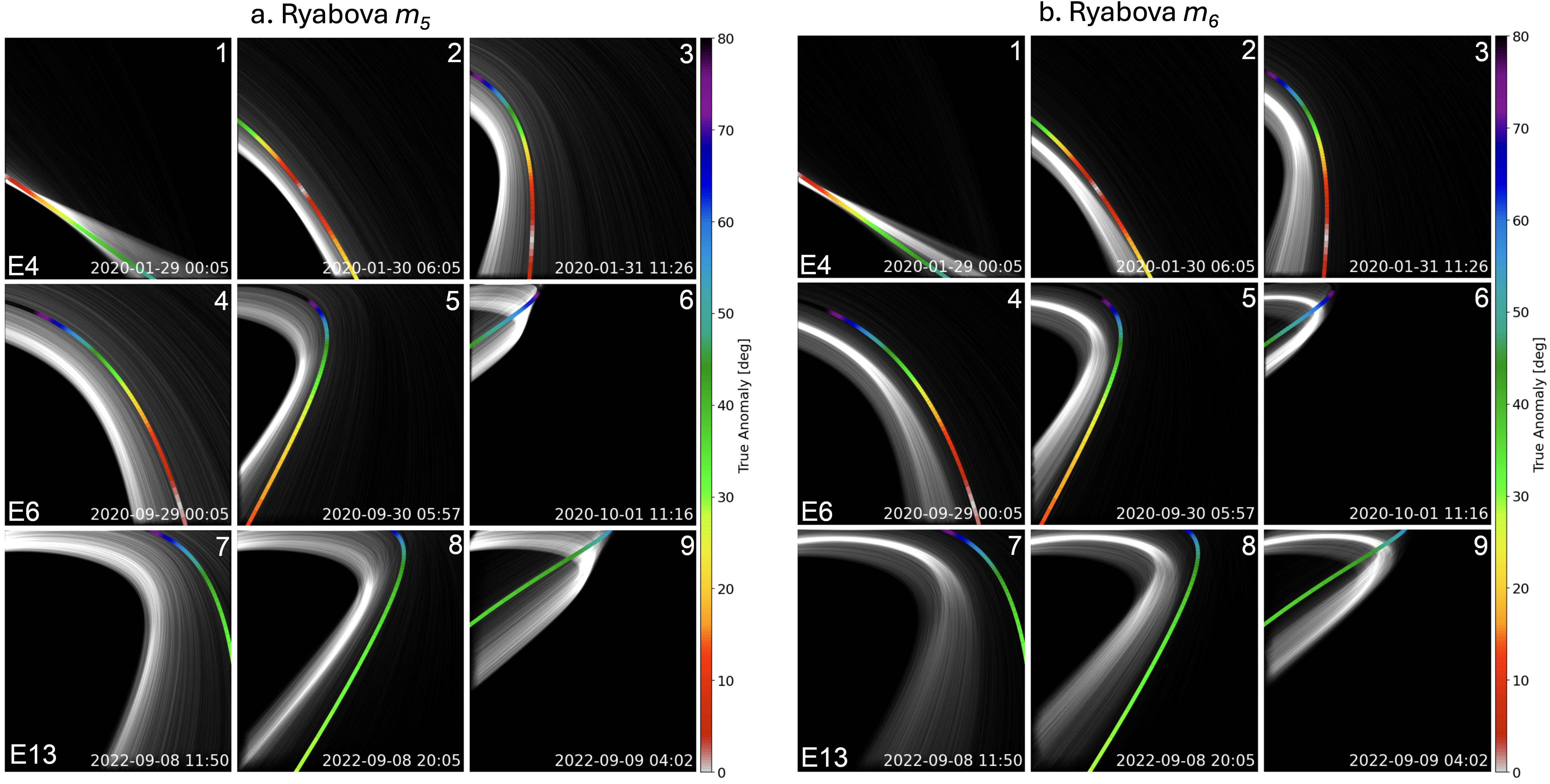}

\caption{Overview of the WISPR-O camera projection of the Ryabova models $m_{5}$ (panel a, left), and $m_{6}$ (panel b, right) into the WISPR FOV during three encounters. Each row represents a different encounter: top row = E4 (1--3), middle = E6 (4--6), bottom = E13 (7--9), with each panel on that row representing a snapshot at a different times in that Encounter, as indicated. These all correspond to the fields of view projected in Figure~\ref{fig:orbit_sims}. The color bar represents Phaethon's current orbit, scaled by TA (degrees). }
\label{fig:ryabova_m5_m6_3x3}
\end{figure}

Figure~\ref{fig:ryabova_m5_m6_3x3} shows snapshots of WISPR-O simulations for the Ryabova $m_{5}$ and $m_{6}$ models respectively, at the times and spacecraft positions indicated in Figure~\ref{fig:orbit_sims}. As discussed in previous publications, the dust trail observed by WISPR always follows extremely closely to Phaethon's orbit, only lying slightly anti-sunward (``right'') of Phaethon's orbit at certain times, e.g., Figure~\ref{fig:phaethon_dust_image}. Thus, for the purposes of this investigation, we will always assume the \textit{actual} location of the dust trail to be as indicated by the color bar, or at most slightly to the right of the color bar.

Visually, $m_{5}$ and $m_{6}$ models are unsurprisingly similar with regard to their positioning and shape. Both models also place the central density of the Geminids inside of Phaethon's orbit, though as discussed in Section~\ref{sec:2.3.3_dist_in_space}, this is a feature common to most Geminid models, and a consequence of the PR effect. The fact that the apparent actual stream core (i.e. the WISPR dust trail) is further from the Sun than as indicated by most models, implies that assumptions of Phaethon's pre-Geminid orbit are likely a key factor in discrepancies between models and observations, as well as the complex nature of the formation mechanism -- something we discuss later in Section~\ref{sec:discuss}. 

One noticeable difference between the $m_{5}$ and $m_{6}$ models is the smaller spread of particles in $m_{6}$ versus $m_{5}$, resulting in a more dense and compressed trail for $m_{6}$. Section~\ref{sec:2.3.3_dist_in_space} explains this effect, illustrated by Figures~\ref{fig:2} and \ref{fig:3}. The relatively higher apparent density of $m_{6}$ implies that we might expect $m_{6}$ dust to be more visible than $m_{5}$ in the WISPR observations. However, viewing geometry (line-of-sight optical depth) of the trail is clearly a complicating factor, particularly when the scene from WISPR's rapidly changing location in space is so dynamic over a matter of hours. 

\subsubsection{CS Model Simulations}

\begin{figure}[ht!]
\centering
\includegraphics[width=0.99\textwidth]{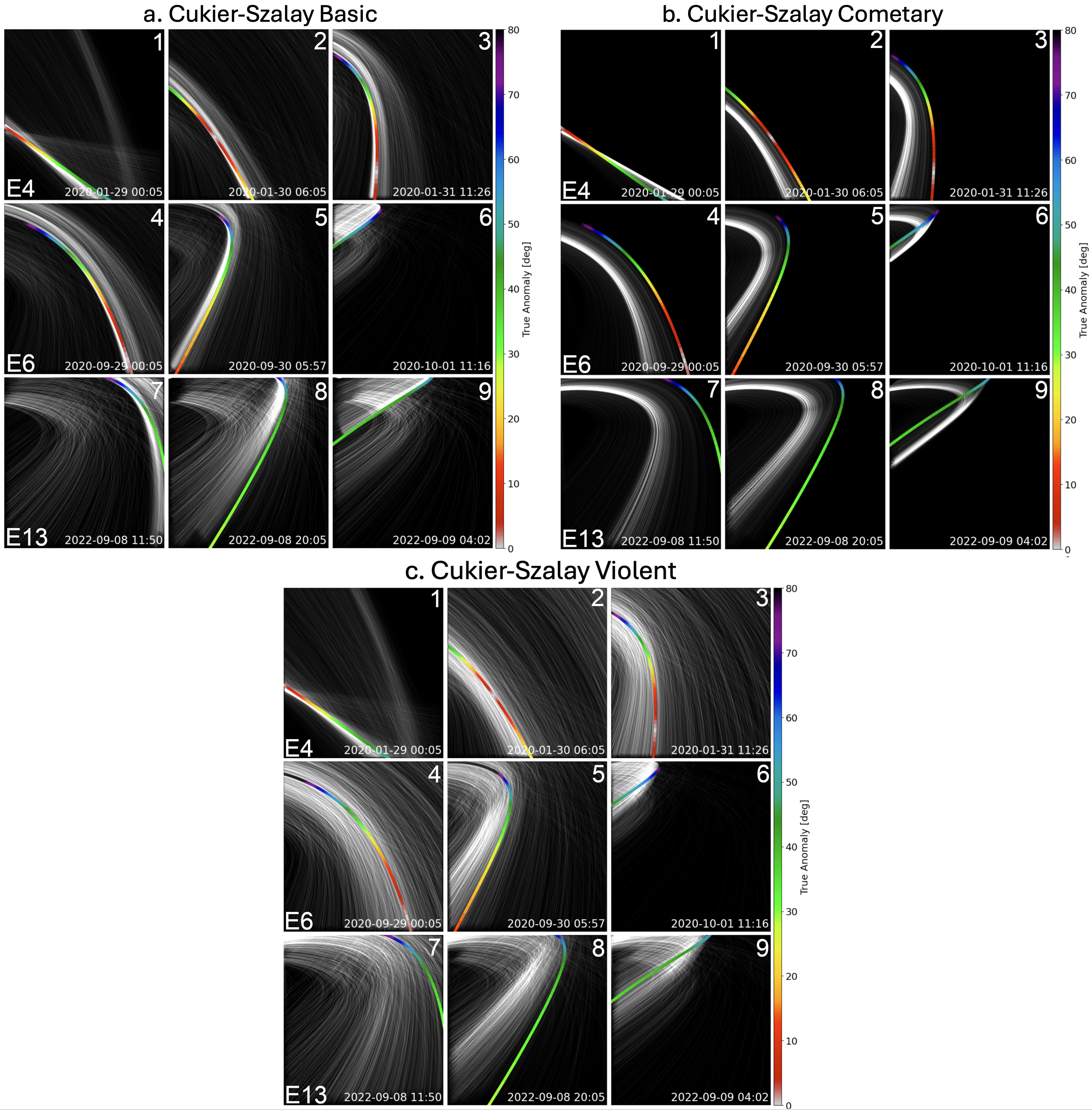}

\caption{Overview of the projection of the CS Basic (Panel a, top-left), Cometary (Panel b, top-right), and Violent (Panel c, bottom) models into the WISPR FOV during three encounters. The formatting of this Figure is the same as described for Figure \ref{fig:ryabova_m5_m6_3x3}}
\label{fig:szalay_all_3x3}
\end{figure}

In Figure~\ref{fig:szalay_all_3x3} we present the three CS models: CS basic, CS cometary, and CS violent. There are several notable features.

The Basic model appears to align closest with the Phaethon trail (and thus WISPR dust trail) versus both Ryabova models, and the Cometary and Violent models. In portions of these simulations, particularly the E4 simulations (Figure~\ref{fig:szalay_all_3x3}(a1 -- a3)), the location of the apparent ``core'' of that model is a close match to the WISPR observations (and is actually hidden by the colorbar in Figure~\ref{fig:szalay_all_3x3}(a1)). This is not surprising for a zero velocity ejection mechanism. However, later Encounters that place the spacecraft closer to the trail visually show that the simulated stream core always remains inside of Phaethon's orbit. 

CS Basic is somewhat clustered into a number of individual filaments. For example, Figure~\ref{fig:szalay_all_3x3}(a1) shows the main core following Phaethon's orbit, and a much fainter secondary filament almost vertical in the frame that results from the assumed size distribution. This particular feature is outside of the FOV in later Encounter simulations. Other filaments can be seen in Figure~\ref{fig:szalay_all_3x3}(a4) where a bright filament aligns closely with the Phaethon orbit, another is slightly anti-sunward, and a diffuse third filament much further out. The physical separation between these filaments is not visually well-resolved until the later Encounters. We discuss some of these individual filaments further in Section~\ref{sec:discuss_obs_prop}. For the purposes of this study, all numerical trend parameters stated for CS Basic will refer exclusively to the bright, narrow core that aligns closest to the Phaethon orbit, and excludes other filaments. The rationale for this is explored in Appendix~\ref{app:A}.

Figure~\ref{fig:szalay_all_3x3} again demonstrates that PSP's increased proximity to the trail provides quite different views, highlighting once more the complexity of analyzing WISPR's observation of this dynamic scene. The CS Basic model in particular becomes substantially broader and fainter in the later encounters, and the per-pixel signal intensity much lower. This mirrors the observation we make in Section~\ref{sec:analysis-psp-update} that the WISPR data themselves became difficult to work with in the most recent Encounters as noise dominated the trail.

The CS Cometary model (Figure~\ref{fig:szalay_all_3x3}b) is \textit{morphologically} the most similar in appearance to the trail, but does not provide a good match to Phaethon's orbit or the WISPR-observed Geminids location. Throughout all Encounters, this model remains largely compact and condensed. While individual filaments can again be seen in the simulations, these are more a function of the number of particles in the simulation than they are a true facet of the model. Adjusting the initial conditions of CS Cometary for a different pre-Geminid orbit could conceivably return an accurate representation of the WISPR scene, albeit with a simplified model. 

The Violent creation model (Figure~\ref{fig:szalay_all_3x3}c) is unsurprisingly broad and diffuse, with widely spread particle streams, though is structurally similar to the Basic model. In the E4 (top row) there is some indication of a stream ``core'', but this becomes lost as PSP's orbit approached the stream in later Encounters. Like the Basic model, a faint (almost vertical) secondary filament can be seen in Figure~\ref{fig:szalay_all_3x3}(c1), but is again lost in later Encounters as WISPR's FOV no longer encompassed the feature. Unlike the basic model, there are no obvious stream clusters in this model, though this could be a consequence of the lower count of particles used in the simulation. 

\subsubsection{Ground Video Observation Simulations}

\begin{figure}[ht!]
\centering
\includegraphics[width=0.99\textwidth]{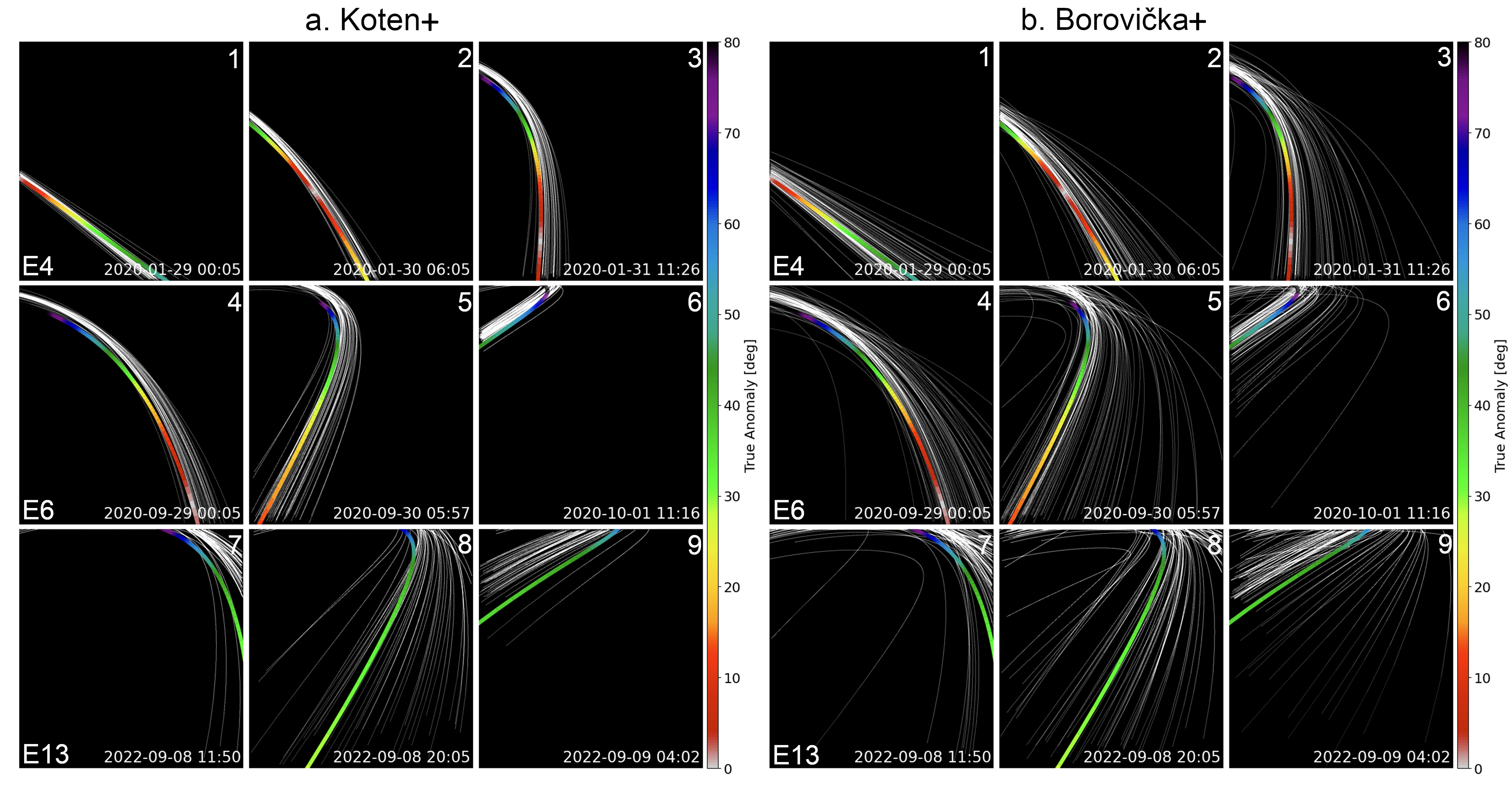}

\caption{Overview of the projection of the Koten et al. (panel a.) and Borovi{\v{c}}ka et al. (panel b.) 
 orbital samples into the WISPR FOV during three encounters. The formatting of the panels is the same as described for Figures \ref{fig:ryabova_m5_m6_3x3} and \ref{fig:szalay_all_3x3}}.
\label{fig:borov_koten_3x3}
\end{figure}

For completeness, Figure~\ref{fig:borov_koten_3x3} presents WISPR simulations of the Koten~et~al. (Figure~\ref{fig:borov_koten_3x3}a) and Borovi{\v{c}}ka~et~al. (Figure~\ref{fig:borov_koten_3x3}b) orbit samples. As detailed in Section~\ref{sec:data_ground_obs}, these are orbits based upon direct observation of individual Geminid meteors. By definition, these particles are Earth-encountering, and thus caution must be exercised in directly comparing them to the other models presented here. The sparsity of observed particles hinders a robust visual comparison with the Ryabova and CS models. However, we can still make some general remarks. 

It is clear from both data sets that the bulk of these observations lie outside of Phaethon's orbit. This is supported by Figures~\ref{fig:7}, \ref{fig:8} and \ref{fig:9}. The low particle density makes it challenging to confidently infer the presence of a ``core'' in the same context as we have noted for other models. Furthermore, the Ryabova and CS models we consider here do not predict the trail core to intersect with Earth, so it is not entirely fair to place Koten~et~al. and Borovi{\v{c}}ka~et~al. particles on the same footing as these models. The latter represent, at best, a tiny fraction of the former. It is, however, interesting to note that despite all being Earth-encountering particles, their apparent distribution across the WISPR FOV becomes quite broad, particularly in later Encounters when the spacecraft was relatively close to these particle orbits. This is in spite of Earth's orbit being a relatively small sampling area. Of course, part of this could be attributed to uncertainties in the calculated particle orbits, and thus again we should not over-interpret this.

We will now proceed to explore the trends in apparent widths of the ``core'' of the models, and the relative offset between Phaethon's orbit and the model cores, both as a function of TA.

\subsection{Models versus Data: Trends Comparison}\label{sec:quant-comp}
\label{sec:results-trends}

\subsubsection{Model/Trail Offset}\label{sec:trail-offset-analysis}

\cite{Battams2022} noted a trend in the apparent offset between the WISPR-observed dust trail and the orbit of Phaethon, as shown in Figure 6 of that study, which showed a steadily increasing distance between them as a function of TA. Here, we now explore the same trend for the Geminid models, using the methodology described in Section~\ref{sec:methods-trends}. The CS Violent models, and ground video observations, were not included here as the spread of particles precluded any reliable ``core'' measurements.

\begin{figure}[ht!]
\centering
\includegraphics[width=0.95\textwidth]{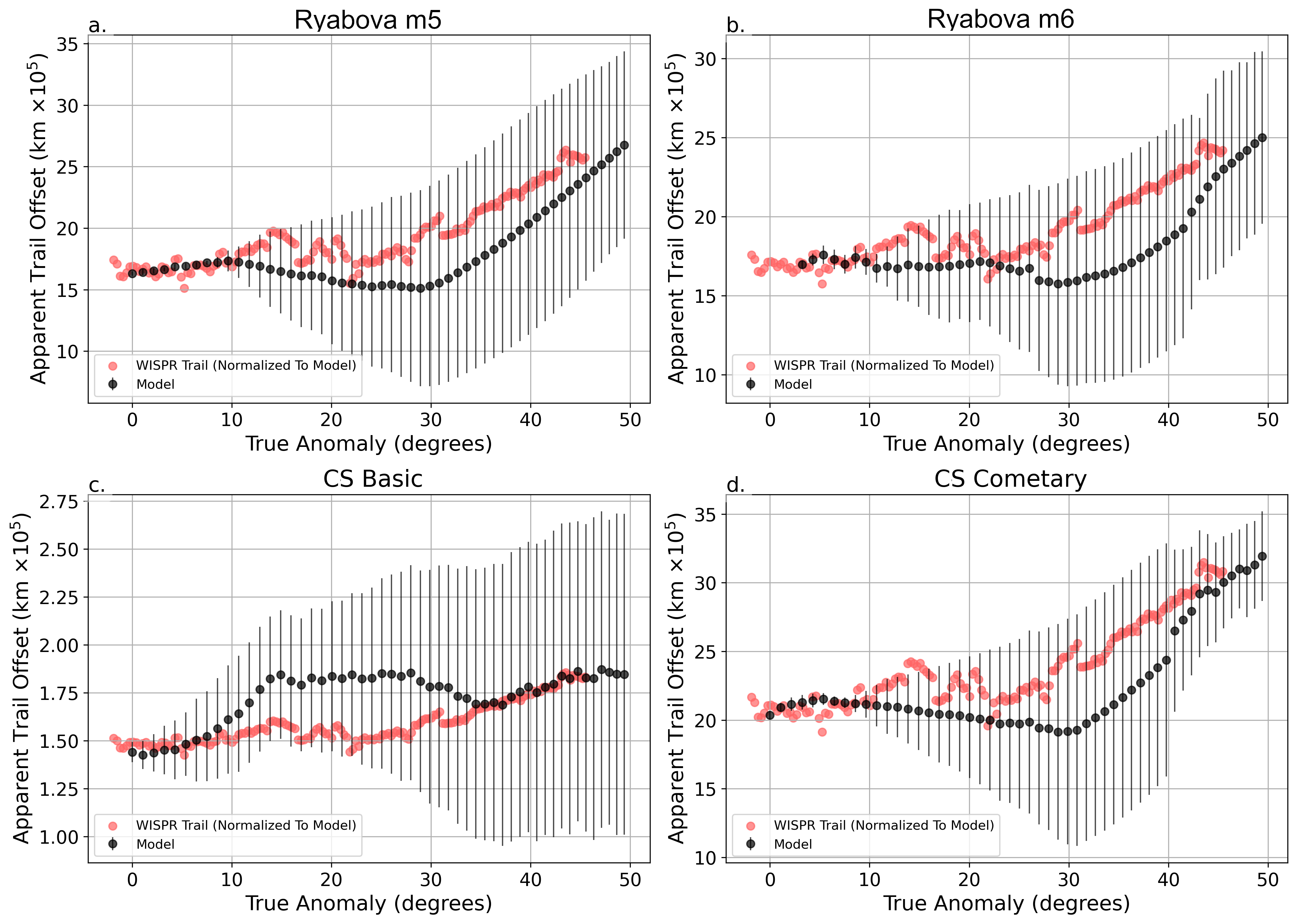}
\caption{Apparent offset of the Geminid models as a function of TA (degrees). WISPR offset values from E6 are also shown (red), with E6 being one of the most reliable data sets for viewing the trail. WISPR offsets are normalized to the models. Offset values provided have been converted to \textit{absolute} values, as all models are inside of Phaethon's orbit.}
\label{fig:offset_plot}
\end{figure}

Figure~\ref{fig:offset_plot} shows the apparent offset between Phaethon's orbit and the core of four Geminid models (red), as a function of Phaethon's TA (black). Here, the ``apparent offset'' is the linear distance in the image plane from Phaethon's orbit to the center of the model core, with that distance corrected for the angular size of a WISPR pixel at Phaethon's orbit as seen from the PSP's location. Offset observations derived from the actual WISPR trail are also shown. As the WISPR dust trail is a factor of 10--20 closer to Phaethon's orbit than most of these models, these WISPR offset distances have been normalized to the model offset values so that they can be compared relatively to each other. Thus, the model observations can be interpreted literally (per the y-axis), but the WISPR observations cannot; they should be considered only in the context of their relative trend as a function of TA. 

It is crucial to point out here that all models remained \textit{interior} to Phaethon's orbit, versus the actual WISPR trail which is marginally \textit{exterior}. Thus, the actual offset values trended opposite to the WISPR observations. Therefore, Figure~\ref{fig:offset_plot} presents the \textit{absolute} values of the offsets, to enable a better comparison with normalized WISPR observations. 

Generally speaking, all models follow a trend similar to the WISPR dust trail; that is, a steady increase as a function of TA. The Ryabova $m_{5}$ and $m_{6}$ models (Fig~\ref{fig:offset_plot}a and \ref{fig:offset_plot}b) both show offset minima $\sim$30\textdegree, which is close to an apparent local minimum in the WISPR observations. This is followed by an increase that mirrors that of the dust trail. At smaller TA values, $m_{5}$ and $m_{6}$ offset values are flatter but still plausibly consistent with WISPR observations.

The CS Cometary model (Figure~\ref{fig:offset_plot}d) is likewise similar to the Ryabova models, with a local minimum $\sim$30\textdegree, and a steady increase after that. At lower TA values, the model again is reasonably flat, though increases slightly to a local maximum $\sim$5\textdegree, similar to that seen in the Ryabova $m_{5}$ slope at $\sim$30\textdegree. The WISPR data themselves show a possible local maximum around 14\textdegree~but we caution that, for reasons discussed in \cite{Battams2022}, the observations in this TA range are particularly challenging to analyze, and the error bars (not shown here) substantially larger. We also note that the apparent physical distances of $m_{5}$, $m_{6}$, and CS Cometary, are highly consistent with each other at $\sim$1.5 -- 2$\times$10$^{6}$ km.

The CS Basic model (Figure~\ref{fig:offset_plot}c) is somewhat different. This was the only model that maintained a core filament very close to the orbit of Phaethon, albeit still slightly inside of that orbit. This is reflected in the scale of the y-axis, which is a factor of ten smaller than the other models. Overall, this model follows the trend of the WISPR observations reasonably well, though deviates noticeably at TA values of $\sim$10\textdegree~to 35\textdegree, and then a somewhat flat profile thereafter. The error bars for this model are particularly large, however. This is due to the relative proximity of a secondary filament, and additional faint filaments, that presented very different line-of-sight effects depending on the observer location. These effects often confused our algorithmic attempts to fit a Gaussian to the cross-trail profiles. Again, we refer to Figure~\ref{fig:szalay_all_3x3} for the corresponding visual summary and Appendix~\ref{app:A} for details on the secondary filament.

\subsubsection{WISPR Trail Width}\label{sec:trail-width-analysis}

\begin{figure}[ht!]
\centering
\includegraphics[width=0.6\textwidth]{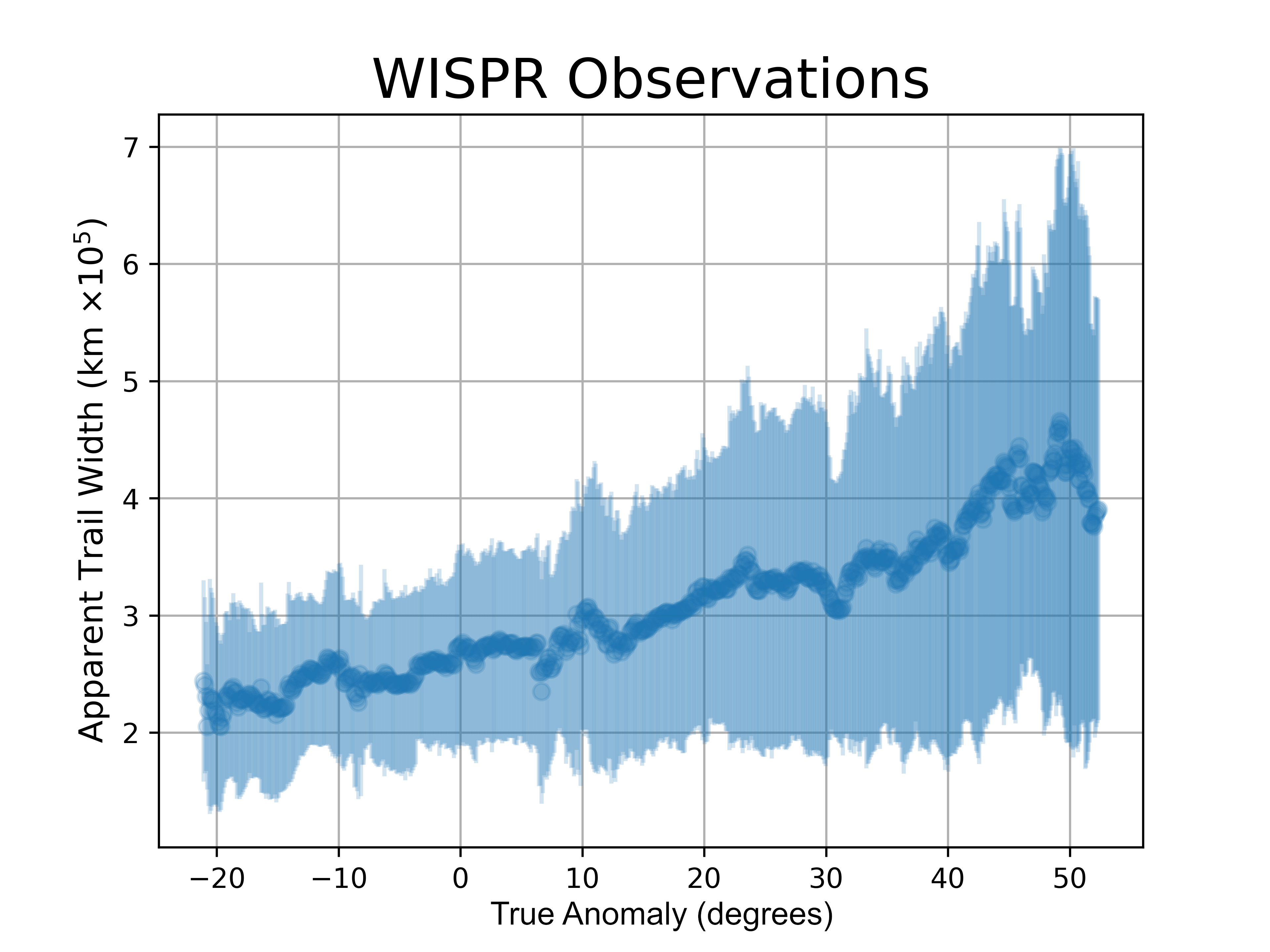}
\caption{Apparent width of the dust trail observed in WISPR E4 through E6 observations, as a function of TA (degrees).}
\label{fig:wispr_width}
\end{figure}

Via the approach described in Appendix~\ref{app:A}, Figure~\ref{fig:wispr_width} presents the observed apparent trail width, using observations combined from E4, E5, and E6 (which provided the most reliable measurements from all Encounters). The solid central blue line is the actual observation value at that point, and the blue bars an indication of the uncertainty of that observation. This uncertainty is based upon the statistics of all observations recorded at each TA angle at different observation times, with viewing geometry being the primary driver of the large variations observed. The width values are intentionally labeled ``\textit{apparent}'', as opposed to ``\textit{actual}'', as our results only represent the portion of the dust trail that is detectable above the noise floor of the observations. In other words, the trail is undoubtedly wider than noted here, but WISPR is unable to detect it. Therefore we will focus primarily on the trends as a function of TA, and less so on the physical dimensions of the trail.

Broadly speaking, we can see that the trail width appears to increase as a function of TA as expected and predicted (e.g., Figure~\ref{fig:4} and related discussion). It is interesting to note that the lowest values appear to be prior to Phaethon's perihelion ($\upsilon = 0$), at TA values of $\upsilon \sim -10^\circ$~to $-20^\circ$. This is somewhat coincident with Figure 6 of \cite{Battams2022}, where we observed that the physical offset between the dust trail and Phaethon's orbit also occurred prior to Phaethon's perihelion at these same approximate values. It is also seen directly in Figure~\ref{fig:4}. While we cannot entirely rule out this being an artifact of the observations or methodology, we can find no obvious procedural source that could create this result, and thus are quite confident it is a true feature of the trail. 

\subsubsection{Model Trail Width}\label{sec:model-trail-width-analysis}

\begin{figure}[ht!]
\centering
\includegraphics[width=0.9\textwidth]{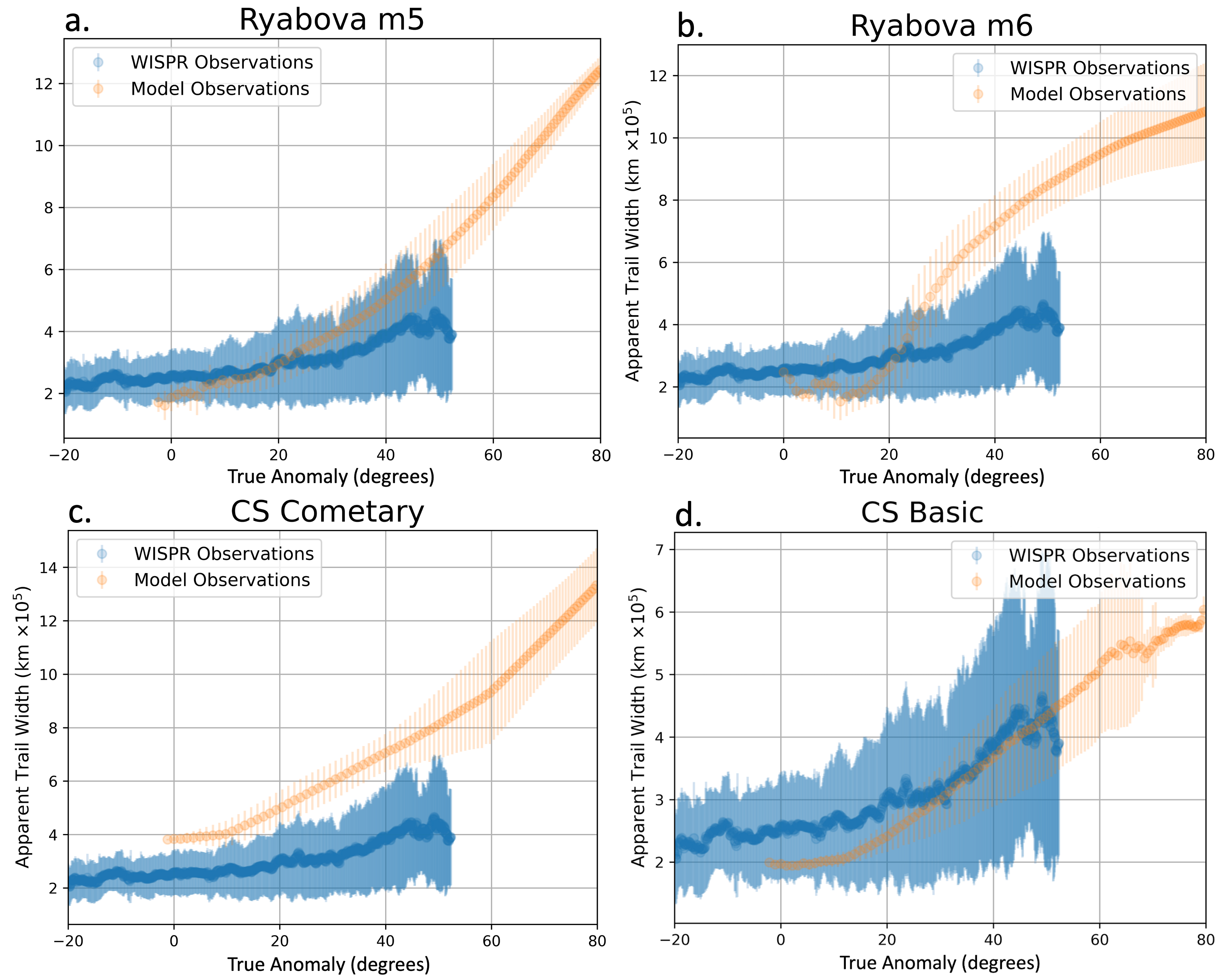}
\caption{Apparent width of the core of the Ryabova $m_{6}$ (a) and $m_{5}$ (b) models, as well as the CS Cometary (c) and Basic (d) models, as a function of TA (degrees). Actual dust trail observations, as seen in Figure~\ref{fig:wispr_width}, are also shown in each panel.}
\label{fig:model_widths}
\end{figure}

Again using the techniques discussed in Appendix~\ref{app:A} and Section~\ref{sec:trail-offset-analysis}, we then looked at the trends in the apparent widths of the models as a function of TA. The CS Violent and ground-based observations were omitted. The results are shown in Figure~\ref{fig:model_widths}.

All four models show the trend of an increasing trail width as a function of TA. This is fully expected behavior for a meteoroid stream. However, each model does present different slope profiles. The Ryabova $m_{6}$ (Fig~\ref{fig:model_widths}a) shows a minimum offset around 10\textdegree, with some possible fluctuations, followed by an increase that follows a steep gradient to $\sim$35\textdegree, and then a lesser gradient thereafter. The $m_{5}$ model (Fig~\ref{fig:model_widths}a) shows almost the opposite gradient, beginning with a shallower slope and increasing as a function of TA. The $m_{5}$ model also has a minimum closer to 0\textdegree~of TA. 

The two CS models (Cometary, Fig~\ref{fig:model_widths}c; and Basic, Fig~\ref{fig:model_widths}d), have approximately similar slopes. The Cometary model in particular presented a very compact and obvious trail ''core'' and thus was substantially easier to capture with our techniques than all other models, hence the very linear/smooth nature of that plot. These results imply minima close to 0\textdegree~TA, although in the absence of model data much below 0\textdegree~we cannot confidently say what the model behavior would be. 

It is interesting to note that our process for arbitrarily adding a noise floor to the simulations resulted in trail core widths comparable to the widths determined from the observations. In particular, the apparent width of the CS Basic model was almost identical to that of the WISPR observations. Once more, we reiterate that these results should be treated with caution, but at the very least they indicate that the modeling approach is providing plausible representations of the actual core of the Geminids.

\section{Discussion} \label{sec:discuss}

\subsection{Observational Properties}\label{sec:discuss_obs_prop} 

A key goal of this study was to determine whether or not the WISPR dust trail is indeed the Geminids. Assuming the trail is indeed a meteoroid stream, it is crucial to understand the general structure of meteoroid trails (using models) to ensure that the observed properties of the WISPR trail do not contradict the basic principles of meteoroid stream behavior. We have three measurable parameters for this: 1) the location of the trail in the WISPR FOV; 2) the trail offset from Phaethon's orbit as a function of TA; and 3) the apparent trail width as a function of the TA. To understand the general structure and behavior of the Geminid stream, we use existing models and our knowledge about physical properties of meteoroids. All three parameters are available for most models presented here.

\cite{Battams2022} analyzed the positions of Geminid particles observed at the Earth (CAMS video observations), noting that they did not reconcile well with the position of the trail observed by WISPR. Thus, it was hypothesized that the WISPR trail could be a separate structure from the Geminids. However, the limitation of ground-based observations is that the errors of orbital parameters in the statistically rich CAMS data are too large for such comparisons to be useful, which is ultimately what the \cite{Battams2022} result demonstrated. Here, we undertook new studies using more precise (but statistically poor) ground-based samples, comparing them both with WISPR observations and with models. These samples lie in the WISPR FOV (Figure~\ref{fig:borov_koten_3x3}) but, due to their sparsity, this was all the useful information we could obtain. As we concluded in Section~\ref{sec:res_ground_dist_in_space}, Earth-encountering particles represent a very small portion of the stream, located probably in its periphery. Thus, the Earth-encountering Geminids probably are a poor proxy for the bulk of the Geminid stream in the context of evaluating the WISPR observations, demonstrating the importance of well-evolved Geminid models for such comparative studies. 

The next step in our approach was to explore visually the extent to which the Geminid models appear in the WISPR FOV and whether they may manifest themselves as a single dense core-like structure. Cross-section plots (e.g., Figures~\ref{fig:2}, \ref{fig:3}, and \ref{fig:6}) show that the models predict the Geminid stream to extend 1 au or more from Phaethon's orbit. However, the bulk of the dust remains slightly interior to the parent orbit due to PR and SRP effects. These models therefore support the expectation of a dense, narrow ``core'' close to Phaethon's orbit. That said, defining the ``core'' is challenging, and our interpretation is somewhat sensitive to how this core is defined. For example, many of the models predict multiple filaments. Treating these as separate streams, versus collectively as a single stream, would affect some of the analyses we perform. 

Simulated observations of these models demonstrate the bulk of Geminid mass to be comfortably within the WISPR field view (Figures~\ref{fig:ryabova_m5_m6_3x3} and \ref{fig:szalay_all_3x3}), with even the sparse Earth-encountering Geminids crossing that FOV (Figure~\ref{fig:borov_koten_3x3}). With the application of noise to these simulations (Appendix~\ref{app:A}), most model simulations reduced to a relatively dense, narrow core of similar \textit{apparent} physical size to the WISPR-observed trail. The one exception was the CS Violent model (Figure~\ref{fig:szalay_all_3x3}c) which, by its nature, presents a more explosive Geminid creation scenario, resulting in widely dispersed particles that do not reconcile with WISPR observations. Exploration of other similarly violent Geminid models (\textit{Quanzhi Ye, Priv. Comm., 2023; not explored here}) typically present this same finding. As noted in Section~\ref{sec:ryabova_ref_pcle}, however, the Ryabova model finds velocities of up to 1.5~km~s$^{-1}$ for $m_{5}$, and 2.2~km~s$^{-1}$ for $m_{6}$, which CS Violent ejection allowed for as well. Thus, high-velocity ejections do not necessarily preclude the existence of a trail core, but such an ejection likely cannot occur instantaneously (e.g., from an impact, or localized outburst). 

We also note that some models, particularly CS models, also predicted additional fainter stream `filaments'. These filaments were representative of specific dust populations of certain masses, often lying far outside of the Phaethon's orbit and either very close or distant to the spacecraft. These filaments are of substantially lower population than the trail core, which is dominated by low mass particles. However, due to the equal weighting our simulation gave to all particles, these filaments were visible in certain viewing geometries prior to the addition of noise to the simulations. There is observational evidence for multiple filaments in the Geminid structure (e.g., \citealt{Szalay2018}), so their presence is not unreasonable, but their visual detection would be far beyond WISPR's imaging capabilities.

As discussed, the meteoroid models generally predict the stream cores to be \textit{interior to the parent orbit}. However, WISPR observes the dust trail to be marginally \textit{exterior to Phaethon's orbit}. Thus we believe that all models presented here share an incorrect assumption: namely, that Phaethon still follows the parent orbit. The Geminid mass is estimated to fall in the range of 10$^{12}$~kg -- 10$^{15}$~kg \citep[e.g.,][]{2017P&SS..143..125R}, representing a substantial portion of the mass of Phaethon ($\sim$10$^{14}$~kg). It therefore seems more appropriate to consider Phaethon and the Geminids as siblings, versus parent-child; a point made directly by \cite{Jenniskens2008a} with related discussion in \cite{Jenniskens2006, Jenniskens2008b}. The release of the Geminids was therefore likely a very dynamic event that saw the parent object potentially shed over half of its mass. Even small mass loss events from comets can result in substantial non-gravitational forces on the parent \citep[e.g.,][]{Sekanina2015, Rafikov2018, Mottola2020}, so the Geminid event, regardless of the mechanism or timescale, likely induced at least a modest (if not substantial) change in Phaethon's orbit. This is not a new concept; the idea that the Phaethon could have transferred to its present orbit from the parent orbit was first proposed by \cite{1985AVest..19..152L}. Given that the WISPR dust trail appears very close to, or just exterior to, Phaethon's present orbit, we could speculate that Phaethon's pre-Geminid orbit was further out than its current orbit, i.e., with a larger semi-major axis. Such a scenario would allow for PR effect forces to place the stream core slightly outside of Phaethon's current orbit, where WISPR observes it to be. 

In this sense, the WISPR observations provide a long sought-after space observation that provides a clear reference point regarding the Geminid stream positioning. While outside of the scope of this study, future consideration should also be given to comparing model \textit{showers} to ground observations, the latter constituting the only other direct reference point we currently have for the system. In the relatively near future, the JAXA/DESTINY+ mission is planned to rendezvous with (3200) Phaethon \citep{sarli:18a}. Onboard it carries the DESTINY+ Dust Analyzer (DDA), a dust detector capable of compositional determination of impacting dust grains with an effective area of 350 cm$^2$ \citep{kobayashi:18a,simolka:24a}. In addition to direct observations of grains from Phaethon's impact ejecta cloud \citep{szalay:19a}, which will provide direct composition measurements of Phaethon and by extension the Geminids, DESTINY+ will make important observations of interplanetary and interstellar dust during its cruise phase \citep{kruger:19a, Kruger2024}. These observations provide an exciting opportunity to further constrain Phaethon's elusive dust complex.

\subsection{Properties of Model/Data Trends}\label{sec:discuss-width}

Having demonstrated that Geminid models can plausibly mirror the visual appearance of the WISPR trail, Section~\ref{sec:quant-comp} explored and compared trends of the model cores and the WISPR trail. Specifically, we focused on the apparent distance between the models and the observations, and the apparent width of the models/trail, as a function of TA.  

\subsubsection{Trail/Model Offset Properties}
\label{sec:trail-offset-properties}

In Section~\ref{sec:trail-offset-analysis} we looked at the trend of apparent trail offset as a function of TA, as first reported in \cite{Battams2022}. Specifically, we sought to determine if the apparent offset for the models increases at a similar relative rate as the observations, and how the minima of the offsets compare between model and observation. To some extent we can estimate this parameter simply from the models themselves. Figure~\ref{fig:5}, for example, shows the location of the maximum point of density in the $m_{5}$ and $m_{6}$ models as a function of TA, in both $\eta$ and $\xi$. However, the point of maximum density does not necessarily equate to the point of the apparent center of the WISPR trail. The latter is a function of the broader distribution of visible dust and the line-of-sight density, as well as any physical phase angle or scattering effects that may come into play. 

The offset results show the brightest filament of CS Basic remains very slightly inside Phaethon's orbit, despite often having a sizable fraction of material predicted to be outside the orbit. The original interpretation of this model treated the whole stream as a single entity and focused on a narrower range of TAs (e.g. Figure 7 of \citealt{Cukier2023}). While that interpretation concluded that CS Basic was primarily outside Phaethon's orbit, our detailed comparisons presented here find this conclusion depends on which filament and TA is considered. The inside/outside distinction also depends on WISPR's diverse viewing geometries, whereas Figure 7 of \cite{Cukier2023} uses a top-down view in Ecliptic J2000. We note that a re-evaluation of the fundamental CS Basic model with updated weighting of the contribution from grains with different $\beta$ values (not shown here) places the core of the model -- per the \cite{Cukier2023} methodology -- much closer to Phaethon's orbit than presented in \cite{Cukier2023}, and effectively provides almost overlap in our interpretations. Thus, based on Figures~\ref{fig:6} and \ref{fig:szalay_all_3x3}, we are still confident that brightest filament (defined here as the ``core'') of CS Basic is for most TAs marginally inside of Phaethon's orbit. However, this is somewhat subject to the apparent viewing geometry of this complex three-dimensional system.

Nonetheless, as shown in Figure~\ref{fig:offset_plot}, the WISPR simulations do show the same trend, i.e. the trail steadily diverging from Phaethon's orbit as a function of TA. This trend is complicated by viewing geometry and limitations of our computational approach, however. This is mirrored by the uncertainty in the model values. In the WISPR observations (red data points, Figure~\ref{fig:offset_plot}) we observe a generally increasing trend, but with a noticeable ``bump'' between TA $\sim$10--25\textdegree. While we cannot be certain of the reality of this feature, or of the smaller bumps seen in the model observations in this figure, their presence is strongly supported by theory. 

The three cometary models ($m_{5}$, $m_{6}$, CS Cometary) present profiles very similar to each other, with trends mirroring the observations quite well. The CS Basic model (Figure~\ref{fig:offset_plot}c) deviates from the observed trend for a large range of TA values, though is modeled to be substantially closer to the Phaethon orbit than the various cometary models. The CS Basic model corresponds to disruption with low ejection speeds at a single location in space. It assumes the entire Geminid mass to be formed at one instance with zero ejection velocity, investigating the limiting case where the parent object sheds the Geminid mass in one moment from a catastrophic disruption. All three cometary models present formation scenarios follow a non-instantaneous, cometary-like formation scenario. 

Section~\ref{sec:2.3.3_dist_in_space} discussed the 3D structure of the stream, commenting on the complex curvature of the dust plane pre- and post-perihelion. The substantial error bars on the offset plots reflect this. These uncertainties are based on the statistics of all fitted Gaussians across all time instances for each TA value, and thus certain viewing geometries will place more dust in a given line-of-sight than at other times. This effect almost certainly occurs in the actual WISPR observations, but is difficult to quantify because WISPR only observes the brightest part of the trail. In short, we would \textit{expect} observations of a stream to show complex variations when imaged from multiple viewpoints. Therefore, the trends and uncertainties of the models are supported by the observations. 

\subsubsection{Trail/Model Width Properties}
\label{sec:trail-width-properties}

We have also investigated the apparent width of the observed trail and the model cores. The width of any meteoroid stream trail is expected to increase as a function of TA (Section~\ref{sec:2.3.3_dist_in_space}), and that is clearly the case for the Geminid models, e.g., Figure~\ref{fig:4}. WISPR's observations, shown in Figure~\ref{fig:wispr_width}, certainly reflect this with a clear increasing trend in apparent width as a function of TA, despite numerous small-scale discontinuities (likely due to noise in the observations). These results also seem to imply that the minimum offset value occurs at TA less than 0\textdegree~(i.e., pre-perihelion). In this regard, the WISPR observations perhaps offer a clue about the age of the stream and Phaethon's initial orbit, with the current stream likely now having precessed from the parent orbit (as noted in Section~\ref{sec:2.3.3_dist_in_space}). However, determining the rate of precession is a non-trivial task, beyond the scope of this study. The offset results for the Ryabova $m_5$ and CS models (Figure~\ref{fig:model_widths}a,c,d) also appear like their minima would be below 0\textdegree~of TA. Unfortunately, our methodology for determining the model offset was unable to reliably capture data at TA$\leq$0\textdegree~for the models, so we cannot be certain of how these trends might behave at those lower TA values. 

Despite the differences in the TA and apparent width values, the simulations largely mirror the trends we expect to see. The CS Basic model (Figure~\ref{fig:model_widths}d) in particular showed a very similar slope to the observations, with the Ryabova $m_6$ arguably the least similar. But again we note that CS Basic is a very simplified example, and also that $m_{6}$ is just one single mass population (versus the mass distribution of the CS models). 

\subsection{Limitations}

Noisy observations and a faint dust signal continue to prove an obstacle for analysis of the WISPR dust trail. While the results presented here are self-consistent, and support models (and vice versa), the error bars in all model parameters are substantial. This is a reflection of both the error in fitting Gaussians to cross-trail profiles \citep{Battams2022}, and the rapidly evolving scene as viewed from PSP's perspective. With regards to the latter, simulations of the models showed substantial differences over periods of even a few hours (e.g., panels 7, 8, and 9 of Figure~\ref{fig:szalay_all_3x3}a). Until models are able to better replicate the scene observed by WISPR, it is impossible to determine the extent to which line-of-sight changes may be affecting the trends in trail/model width and offset.

Further to this, we have intentionally ignored any effects of dust grain properties (size, fluffiness, etc.), or phase angle scattering effects. Our simulations are simply line-of-sight integrations. However, it was noted in \cite{Battams2020, Battams2022} that the phase angle of the trail observed by WISPR was not particularly noteworthy in the context of phase angle effects, so this latter point may be quite minor. Also, no substantial photometric (visual magnitude) differences were seen as a function of TA along the trail. Nonetheless, injecting some basic physics into the WISPR simulations may prove fruitful.

Another minor, but non-negligible, point is that the brightness of the WISPR trail is dominated by the smallest particles, but the PR and SRP forces over the relevant timescales place limitations on the minimum size of particle that we expect to still be present. That is, the smallest particles on bound orbits should no longer be present. However, it is certainly possible that small particles exist in the actual stream as a result of collisions. Modeling this would be a substantial undertaking, outside of the scope of this study, and thus we assume that the only particles in these models are those ejected at formation time, following bound orbits.

Finally, for any of the presented models to better capture the trail location, the initial parent orbit cannot be Phaethon's present orbit. This returns us to an important point: all models (understandably) continue to skirt the much larger problem of initial conditions -- that is, Phaethon's pre-Geminid orbit. Simply changing the parent orbit for the cometary mechanisms to the (unknown) ``true'' value could conceivably place these model cores into the observed location. However, the number of degrees of freedom make this a daunting task. Nonetheless, modern computing architectures may now facilitate a more comprehensive exploration of the entire parameter space that encompasses reasonable guesses for the parent orbit. Such an investigation may also narrow the scope on the duration of the Geminid release event, and/or whether non-gravitational (sublimation-related) factors played any role in modifying Phaethon's post-Geminid orbit. It is therefore a broad recommendation of this study that the community consider expanding investigations in this direction. 

\section{Conclusions} \label{sec:concl}

We have demonstrated that the WISPR dust trail is indeed the Geminid stream, exhibiting properties entirely consistent with meteoroid streams in general, and as expected for the Geminids specifically. We have also compared this trail to Geminid models, where each reproduced different features of the stream to varying extents.  We find that: 
\begin{itemize}
    \item Nearly all Geminid models predict the existence of a ``density core'' of the Geminid stream, consistent with the trail as seen by WISPR;
    \item All models manifest similar trends in trail offset and trail width as a function of true anomaly; and
    \item The specific behavior of these trends around perihelion is supported by the 3D structure of the stream models;
    \item Models, where particle ejection/release happens around the entire parent orbit, predict the core of the Geminids to be inside of the parent orbit. In contrast, models with an instantaneous disruption scenario predict that a substantial portion of material lies near or outside the orbit of Phaethon.
\end{itemize}

Only the last item partly contradicts WISPR observations, which show the core slightly external to Phaethon's current orbit. We propose that the Geminid parent orbit was likely substantially different to Phaethon's present orbit, with a larger semi-major axis. Such a scenario could place model cores closer to the observed trail location and provide for more realistic stream widths.

This work represents a new starting point for future Geminid modeling studies. We have demonstrated that the basic principles of meteoroid stream modeling apply well to the Phaethon-Geminid system, and predict aspects of the stream structure consistent with that observed by WISPR. Future efforts may benefit from focusing on initial starting conditions. In particular, consideration should be given to Phaethon's pre-Geminid orbit in order to evolve the Geminid core closer to its observed location. Crucially, we now have two reference points to fit the models: ground and space observations.

To further our research we plan to explore options for stacking data obtained by PSP/WISPR over multiple encounters in an attempt to enhance the trail signature. The PSP science team also continue to monitor spacecraft dust impact rates, some of which have been attributed to beta meteoroids from the Geminids \citep{Malaspina2023, Szalay2021}; it will be interesting to see how these impact rates change with PSP now so close to Phaethon's orbit (and thus presumably the trail).

\begin{acknowledgments}
\begin{center}
    
ACKNOWLEDGMENTS
\end{center}
Parker Solar Probe was designed, built, and is now operated by the Johns Hopkins Applied Physics Laboratory as part of NASA's Living with a Star (LWS) program (contract NNN06AA01C). Support from the LWS management and technical team has played a critical role in the success of the Parker Solar Probe mission. The Wide-Field Imager for Parker Solar Probe (WISPR) instrument was designed, built, and is now operated by the US Naval Research Laboratory in collaboration with Johns Hopkins University/Applied Physics Laboratory, California Institute of Technology/Jet Propulsion Laboratory, University of Gottingen, Germany, Centre Spatiale de Liege, Belgium and University of Toulouse/Research Institute in Astrophysics and Planetology. KB, AL, and BG were supported by the NASA Parker Solar Probe WISPR Project. 
GR was supported by the state assignment of the Ministry of Science and Higher Education of the Russian Federation (theme No. FSWM-2024-0005). J. Szalay acknowledges the Parker Solar Probe Guest Investigator Program, grant 80NSSC21K1764. This research has made use of NASA's Astrophysics Data System. We would like to thank the anonymous referee for helpful feedback and suggestions.
\end{acknowledgments}


\newpage
\appendix

\section{WISPR and Model Data Treatment}\label{app:A}

A number of treatments were required for analysis of both the WISPR observations and the various models under consideration. Fundamentally, our approach for establishing quantitative (i.e., width and offset) parameters for the WISPR data was to employ the techniques described in \cite{Battams2022}. This allowed us to smooth and stack multiple cross-trail profiles from the WISPR observations to improve the signal-to-noise ratio, and enable the fitting of a Gaussian to the resulting profile. The only substantial change from the technique used in \cite{Battams2022} was that here we now sample the trail at more TA values. 

\begin{figure}[ht!]
\centering
\includegraphics[width=0.95\textwidth]{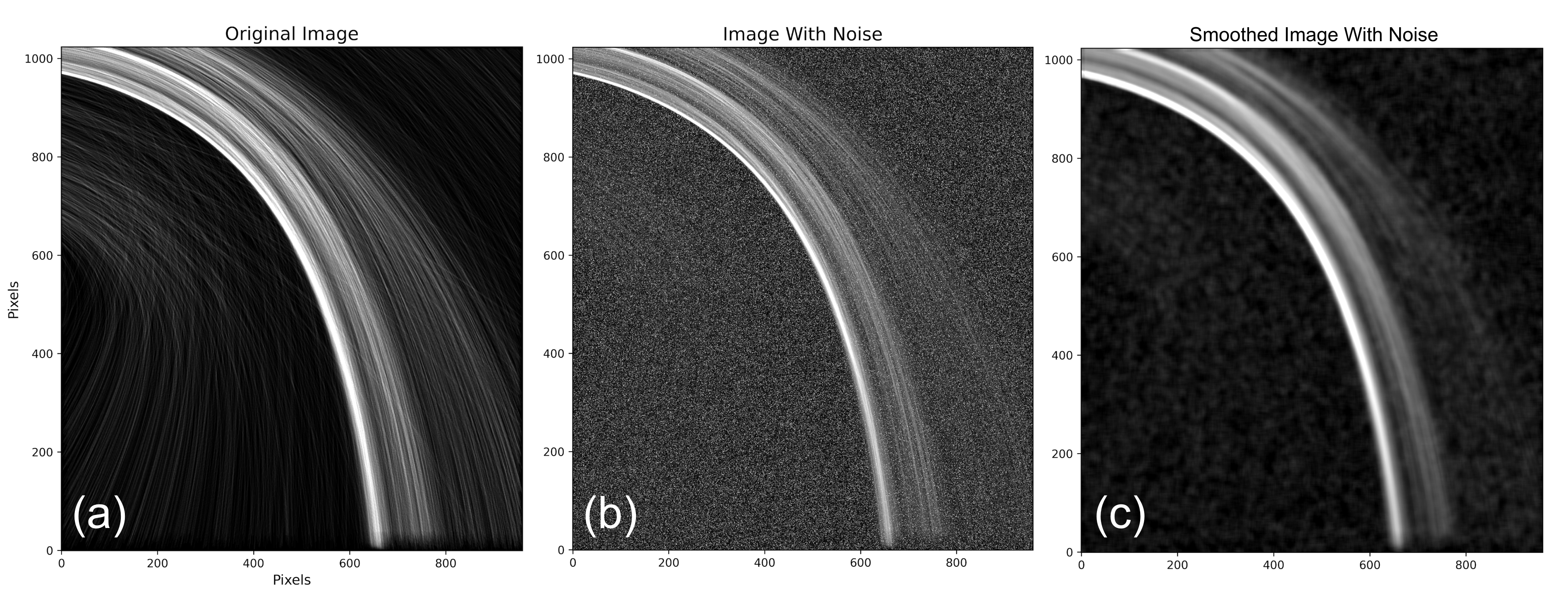}
\caption{Illustration of the treatment of simulation data prior to Gaussian fitting, showing (a) the basic simulation image; (b) the simulation image plus a random noise floor; and (c) the noisy simulation image smoothed. This particular frame is taken from the CS Basic model.}
\label{fig:model_noise}
\end{figure}

The same techniques were also applied to the model data, with some minor modification, as the broad spread of the individual model particles across the FOV presented challenges to the fitting process. We overcame this by applying a random noise floor to the simulated images sufficient to remove most of the faint, individual trail strands (which we would not expect WISPR to detect anyway). The intensity of this noise floor was determined by-eye, with our criteria simply that it removes most of the faint model strands and leaves a bright ''core'' at the densest portion of the trail. An identical noise floor was shared by all models. Figure~\ref{fig:model_noise} uses a single image from the CS Basic model to illustrate the processing applied to the model observations. \textit{Panel a} shows the full simulation frame, without modification. \textit{Panel b} shows the same frame with a uniform random noise floor added to wash out the fainter particle tracks (a radial noise gradient to mimic the corona was also tested, but not found to make a significant difference to the results). \textit{Panel c} shows the smoothed version of the noisy simulation; it was these frames that were used for Gaussian fitting. 

This processing was effective for all but the CS Violent model (Figure~\ref{fig:szalay_all_3x3}c) which was too broadly distributed across the field, and had too few data points, to reduce to a single ''core''. Thus, our analyses focused mainly on the Ryabova $m_{5}$ and m6, and CS Basic and Cometary models. In all other cases, the remaining model ''cores'' were sufficient that our Gaussian-fitting techniques could be applied. 

\begin{figure}[ht!]
\centering
\includegraphics[width=0.95\textwidth]{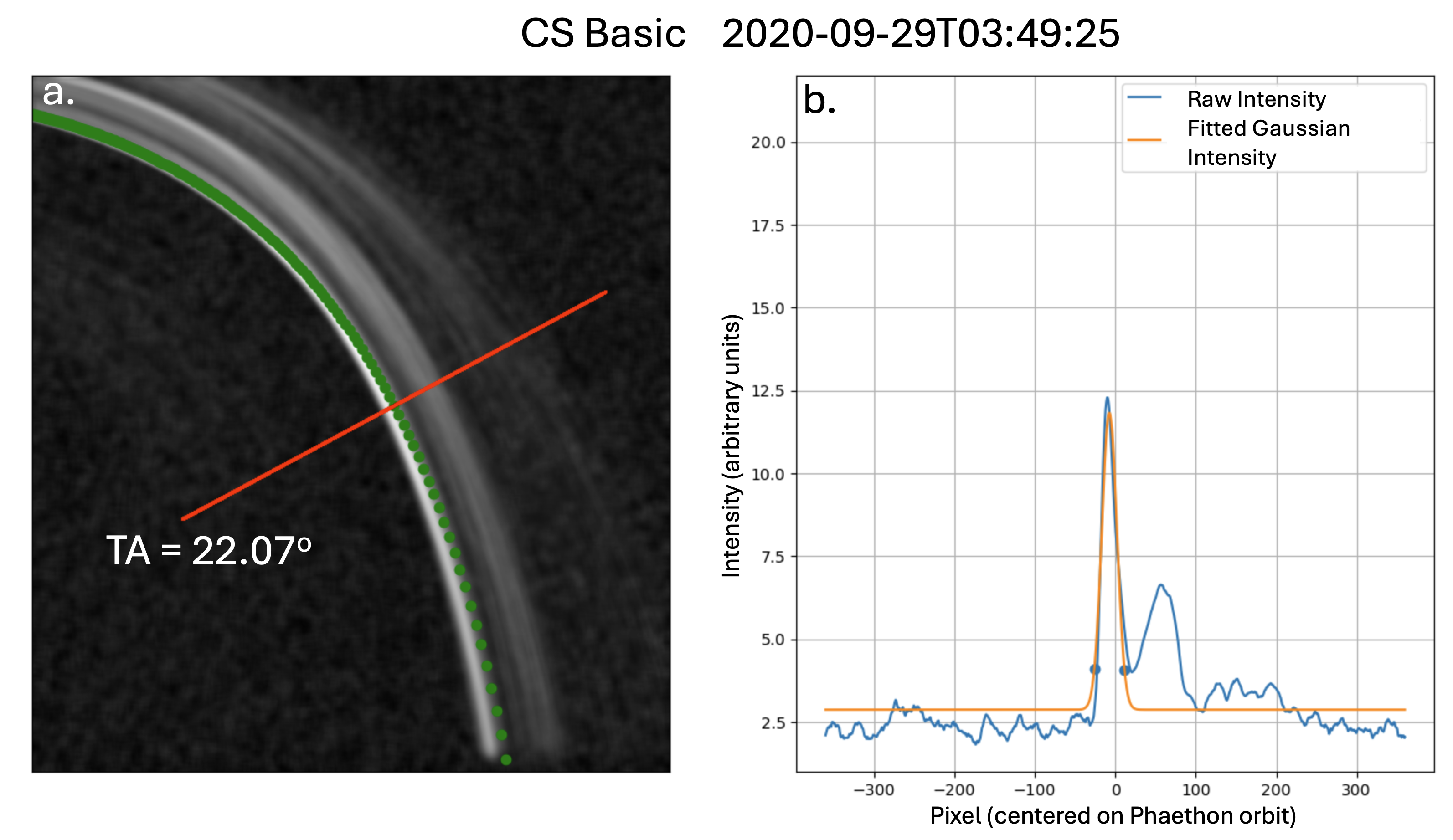}
\caption{Illustration of a cross-trail profile extracted from CS Basic at TA = 22.07\textdegree~(panel a, red line), and the resulting intensity along that line (panel b, blue) and intensity of the Gaussian fit (Panel b, orange).}
\label{fig:csbasic-profile}
\end{figure}

However, an additional adjustment was required for CS Basic. As illustrated in Figure~\ref{fig:model_noise}, this model exhibited multiple filaments of different intensities. Figure~\ref{fig:csbasic-profile} shows a cross-trail profile of this model at an arbitrarily selected TA = 22.07\textdegree~(panel a), and the resulting intensity profile (blue, panel b). Here we can see the primary density core just inside of Phaethon's orbit (pixel = 0, panel b), fitted by the Gaussian (orange, panel b), as well as the much broader secondary filament outside of Phaethon's orbit (pixel = $\sim$25--100, panel b). We also faintly detect a third broad filament farther out (pixel = $\sim$100--200, panel b). As the purpose of this investigation was to study the narrow intensity cores of these models, it was decided to computationally exclude the secondary filament of the CS Basic model, despite it being photometrically apparent in the simulations. The rationale for this is that if WISPR is only detecting one narrow trail, and if we assume the CS Basic model to be an entirely perfect representation of the system, then WISPR can only be seeing that brighter, primary filament.

\newpage

\bibliography{references}{}
\bibliographystyle{aasjournal}
\end{CJK*}
\end{document}